\pdfoutput=1
\documentclass{aa}  

\usepackage{graphicx}
\usepackage{txfonts}
\begin{document}

   \title{Tracing the origins of galaxy lopsidedness across cosmic time}
   \titlerunning{Tracing the origins of galaxy lopsidedness across cosmic time}
   \authorrunning{A. Dolfi et al.}


   \author{Arianna Dolfi,
          \inst{1}
          Facundo A. G\'omez,
          \inst{1}
          Antonela Monachesi,
          \inst{1}
          Patricia B. Tissera,
          \inst{3,4}
          Crist\'obal Sif\'on,
          \inst{5}
          Gaspar Galaz
          \inst{3}
    }

   \institute{Departamento de Astronom\'ia, Universidad de La Serena, Av. Ra\'ul Bitr\'an 1305, La Serena, Chile\\
   \email{arianna.dolfi@userena.cl}
        \and
        Instituto de Astrofísica, Pontificia Universidad Católica de Chile, Av. Vicuña Mackenna 4860, Santiago, Chile
        \and 
        Centro de Astro-Ingenieria, Pontificia Universidad Católica de Chile, Av. Vicuña Mackenna 4860, Santiago, Chile
        \and 
        Instituto de F\'isica, Pontificia Universidad Cat\'olica de Valpara\'iso, Casilla 4059, Valpara\'iso, Chile        
    }

   \date{Received XXX; accepted YYY}

 
  \abstract
   {Current studies of large-scale asymmetries (i.e. lopsidedness) in the stellar density distribution of disk galaxies have mainly focused on the local Universe. However, recent observations have found a significant fraction (over $60\%$) of lopsided galaxies at high-redshift (i.e. $1.5 \lesssim z \lesssim 3$), which is significantly larger than the fraction ($\sim30\%$) observed in the nearby Universe.}
   {We aim to understand whether simulations can reproduce the observed fraction of lopsided galaxies at high-redshift, as well as whether the more widespread lopsidedness at high- than low-redshift can be associated to environmental mechanisms being more effective in producing lopsided perturbations at high-redshift.}
   {At each redshift between $0 < z < 2$, we independently select a sample of disk-like galaxies from the IllustrisTNG simulations. We then characterize lopsidedness in the disks of galaxies at each redshift, study the relevant mechanisms generating lopsidedness, as well as the correlation between such perturbation, the local environment and the galaxy internal properties as a function of redshift.}
   {Consistent with previous and new observational results, we find that: 1) simulations predict a significant fraction ($\sim60\%$) of lopsided galaxies at high-redshift (i.e. $1.5 < z < 2$), 2) the fraction of lopsided galaxies, as well as the lopsided amplitude, decreases from high- to low-redshift, meaning that galaxies become more symmetric towards low-redshift, and 3) there is not a significant dependence of lopsidedness on the local environment, but there is a strong correlation between the lopsided amplitude and basic galaxies' structural properties at all redshifts between $0 < z < 2$. This means that, independent of the mechanisms on-setting lopsidedness, galaxies with low central stellar mass density and more extended disks are more susceptible of developing strong lopsidedness.
   We find that both recent interactions with mass-ratio $>1$:$10$ and gas accretion with subsequent star formation can produce lopsided perturbations at all redshift, but they are both significantly more effective at high-redshift.}
   {These results suggest that the mechanisms behind lopsidedness vary across cosmic time, with a greater influence from environmental interactions and gas accretion at higher redshift.}


   \maketitle
%
\section{Introduction}

It has been known for a long time that the present-day distribution of light and mass of disk galaxies is often non axis-symmetric, with one side of the galaxy being more extended than the opposite side. Such asymmetries, which were first reported in the early work of \citet{Baldwin1980}, are currently known as "lopsidedness". 
The lopsided feature has been detected in galaxies in the spatial distribution of both the old stellar component observed in the near infra-red \citep{Bloch1994,Rix1995} and HI gas \citep{Richter1994,Haynes1998}, as well as in the large-scale kinematics of the HI gas \citep{Swaters1999}. Overall, observational studies have shown that, in the local Universe, at least $30\%$ of disk galaxies have lopsided stellar disks, while $50\%$ of disk galaxies have lopsided HI distribution \citep{Zaritsky1997,Bournaud2005}, suggesting that lopsidedness is a common phenomenon in the disks of present-day galaxies.

Several mechanisms have been proposed to explain the origin of lopsidedness. The most commonly proposed mechanisms typically involve external processes, such as tidal encounters, dynamical friction, mergers and asymmetric gas accretion \citep{Beale1969,Walker1996,Bournaud2005,Kipper2020}, or internal dynamical processes within the disk, such as gravitational instabilities or instabilities in counter-rotating disks \citep{Masset1997,Saha2007}. Additional mechanisms also suggest that lopsidedness can originate from the response of the disk to a previously perturbed or distorted dark matter halo \citep{Weinberg1995,Jog1997}, or from torques acting as a result of an off-centered disk with respect to the dark matter halo \citep{Nordermeer2001}.

Other works have found a strong correlation between the lopsided amplitude and the internal properties of the galaxies \citep{Conselice2000,Reichard2008,Varela-Lavin2023,Dolfi2023}. Specifically, \citet{Varela-Lavin2023} characterized lopsidedness in a sample of present-day Milky Way-mass galaxies (i.e. virial mass $\mathrm{M}_{200}\sim10^{11.5}$-$10^{12.5}\, \mathrm{M_{\odot}}$), using the cosmological hydrodynamical simulations from the IllustrisTNG project (i.e. TNG50; \citealt{Nelson2019}). They found a strong dependence of lopsidedness on the central stellar mass density and tidal force exerted by the inner disk onto its outskirts. Strongly lopsided galaxies were also found to be typically characterized by more extended disks, lower central stellar mass density and less self-gravitating inner galactic regions than symmetric galaxies at $z=0$. This is generally consistent with the observational results by \citet{Conselice2000} and \citet{Reichard2008}, thus suggesting that it is the self-gravitating nature of the inner galactic regions what determines whether a galaxy can develop strong lopsidedness as a result of interactions, mergers or torques exerted by distorted dark matter halos. In a recent work, \citet{Dolfi2023} have extended the analysis of \citet{Varela-Lavin2023} by considering a larger sample of disk-like galaxies in the TNG50 simulation, which includes both centrals and satellites embedded in a range of different environments (i.e. $\mathrm{M}_{200}\sim10^{10.5}$-$10^{14}\, \mathrm{M_{\odot}}$). The aim was to further study the relation between lopsidedness, local environment and internal properties of the galaxies at $z=0$. Regardless of the environmental metric used, \citet{Dolfi2023} found a lack of correlation between lopsidedness and local environment, and a strong correlation between lopsidedness and the galaxy internal properties, independently of the environment. More importantly, \citet{Dolfi2023} studied the star formation histories of their lopsided and symmetric galaxy samples. They found that lopsided galaxies have a very distinct star formation history with respect to their symmetric counterparts. Lopsided galaxies have typically assembled over longer timescales with an overall constant star formation rate up to $z=0$, while symmetric galaxies have typically assembled at early times between $8$-$6\, \mathrm{Gyr\, ago}$ with relatively short and intense burst of central star formation. These results suggest that lopsidedness in present-day disk galaxies is strongly connected to their early assembly histories.

While the previously discussed works have mainly focused on understanding the origin of lopsidedness at $z\sim0$, recent works have detected lopsidedness in galaxies up to $z\sim3$ \citep{Kalita2022,LeBail2023}. Specifically, \citet{LeBail2023} have studied the morphological and physical properties of a small sample of $22$ dusty star-forming galaxies with $\mathrm{M_{*}}\sim10^{10}$-$10^{11.5}\, \mathrm{M_{\odot}}$, observed with the James Webb Space Telescope (JWST) Near Infra-Red Camera between $1.5 \lesssim z \lesssim 3$. 
They have found that a significant fraction of those galaxies (i.e. $64\%$) is lopsided. Therefore, the comparison between the fraction of lopsided galaxies detected at high- ($64\%$; \citealt{LeBail2023}) and low- ($30\%$; \citealt{Zaritsky1997}) redshift suggests that lopsidedness is a more common phenomenon at high-redshift. 
Interestingly, \citet{LeBail2023} have also found evidence of a correlation between lopsidedness and the internal properties of the galaxies at high-redshift. They showed that galaxies with a quenched bulge component (TypeIII galaxies), characterized by high core mass fraction, have typically lower asymmetry than galaxies with a star-forming bulge component (TypeI and TypeII galaxies), which are characterized by low core mass fraction. 
Since TypeIII galaxies in \citet{LeBail2023} are located, on average, at lower redshift and have lower specific star formation rate than TypeI and TypeII galaxies, the authors suggested that galaxies tend to become more symmetric towards low-redshift due to the build up of a massive quenched bulge component that can stabilize the disk against lopsided perturbations. 
Overall, these results also suggest a strong correlation between lopsidedness and the internal properties of the galaxies, as similarly found in previous observational \citep{Conselice2000,Reichard2008} and numerical \citep{Varela-Lavin2023,Dolfi2023} studies at $z=0$. 

Currently, lopsidedness in simulations has been mainly investigated at $z=0$ \citep{Lokas2022,Varela-Lavin2023,Dolfi2023}. 
In this work, we use the IllustrisTNG\footnote{\url{https://www.tng-project.org/}.} cosmological hydrodynamical simulations, in particular the TNG50 run, to characterize lopsidedness in the disks of galaxies up to $z=2$. We study the relevant mechanisms generating lopsidedness, as well as the correlation between lopsidedness, local environment and galaxy internal properties at different redshifts. We aim to understand whether the more widespread lopsidedness at high- than low-redshift may be associated to environmental effects playing a more active role in triggering lopsidedness at high-redshift. In particular, we aim to understand whether the higher merger and interaction rate at high- than low-redshift \citep{Conselice2014} can be responsible for the high fraction of observed lopsided galaxies between $1.5 \lesssim z \lesssim 3$ \citep{LeBail2023}.

The paper is structured as follows. In Sec. \ref{sec:sample_selection}, we briefly describe the simulations and the selection of our sample of disk-like galaxies. In Sec.\ref{sec:lopsidedness}, we describe the method used to quantify the global lopsidedness in our selected galaxy sample. In Sec. \ref{sec:results}, we study the evolution of the fraction of lopsided galaxies, as well as lopsided amplitude, as a function of redshift. We also study the correlation between lopsidedness, local environment and galaxy internal properties as a function of redshift. Finally, we investigate the potential impact of galaxy interactions, as well as gas accretion, as triggering mechanisms of lopsidedness. 
In Sec. \ref{sec:comparison_observations}, we perform a comparison between our results and those obtained in the recent observations of \citet{LeBail2023}. Finally, in Sec. \ref{sec:conclusions}, we provide a summary of our results and conclusions.

\section{The Data}
\label{sec:sample_selection}

\subsection{The IllustrisTNG simulations}
\label{sec:simulations}
The IllustrisTNG simulations are a suite of cosmological gravo-magnetohydrodynamical simulations that include a comprehensive physical model of galaxy formation designed to realistically trace the formation and evolution of galaxies across cosmic time \citep{Vogelsberger2014,Genel2014,Sijacki2015,Nelson2015,Weinberger2017,Pillepich2018}. The IllustrisTNG simulations are run with the moving-mesh code \texttt{Arepo} \citep{Springel2010} and are publicly available. A detailed description of the IllustrisTNG simulations can be found in the release paper by \citet{Nelson2019}. 
In this work, we use the smallest cosmological volume of the simulations (i.e. TNG50: $51.7\, \mathrm{cMpc^{3}}$; \citealt{Weinberger2017,Pillepich2018}), which is characterized by the highest resolution (i.e. initial gas cell mass $8.5\times 10^{4}\, \mathrm{M_\odot}$). The high resolution of TNG50 allows us to better characterize and quantify asymmetries in the stellar mass density distribution of galaxies.

\subsection{Sample selection}
\label{sec:sample_selection}
From the TNG50 simulation, we select all galaxies with total stellar mass between $10^{10}\, \mathrm{M}_{\odot} \leq \mathrm{M}_{*} \leq 10^{11.5}\, \mathrm{M}_{\odot}$ and total number of bound stellar particles $\mathrm{N}_{\mathrm{tot,\, stars}} \geq 10^{4}$ at each redshift between $0 < z < 2$\footnote{We note here that IllustrisTNG contains $20$ "full" and $80$ "mini" snapshots. The "mini" snapshots only have a subset of particle fields available. In this work, we use the "full" snapshots of the TNG50 simulation within redshift range $0 < z < 2$.}. The first selection criterion is chosen to match the stellar mass range of the sample of $22$ dusty star-forming galaxies studied by \citet{LeBail2023}, which were observed using the JWST Near Infra-Red Camera between $1.5 < z < 3$. The second selection criterion is to ensure that each galaxy is well resolved to quantify differences in the stellar mass density distribution, similarly to \citet{Varela-Lavin2023,Dolfi2023}.
We note here that we do not extend the redshift range up to $z=3$ to match the observations from \citet{LeBail2023} due to the low number of disk-like ($<100$) obtained from the TNG50 simulation, as well as to their strongly perturbed morphology, at $z=3$.

\subsection{Defining galaxy sizes}
\label{sec:bulge_and_disk}
For all the selected galaxies in Sec. \ref{sec:sample_selection}, we define the following sizes calculated from the stellar particles alone: half-mass radius $R_{\mathrm{h}}$, disk size $R_{90}$ and disk height $h_{90}$. The stellar half-mass radius and disk size are defined as the radii enclosing $50\%$ and $90\%$ of the total galaxy stellar mass, respectively. These radii are calculated with respect to the position of the galaxy's particle with the minimum gravitational potential energy, which is taken as the galaxy's center. Similarly, the disk height is defined as the vertical distance above and below the disk plane enclosing $90\%$ of the total galaxy stellar mass. Following \citet{Iza2022}, we calculate the disk height along both the positive (i.e. $h_{90,+}$) and negative (i.e. $h_{90,-}$) $z$-directions, considering only stellar particles within the cylindrical region of radius $1\, R_{90}$. Then, we calculate the average value between $h_{90,+}$ and $h_{90,-}$ to define the disk height $h_{90}$ above and below the disk plane. The overall disk width is then defined as twice the disk height, i.e. the vertical region between $-h_{90} < z < +h_{90}$. The calculation of all these quantities is performed after rotating the galaxy in the face-on projection, such that the galaxy disk lies on the $xy$-plane.
We note that the calculation of the disk radii and disk height of the galaxies does not separate between bulge and disk components, and we do not use scale lengths to define the galaxy's sizes as in observations. For the objectives of this work, we are interested in defining quantities that can represent the boundaries of the disk to study its stellar mass distribution without the contamination from external sources or streams in the galactic halo.

Finally, we define the central regions of the galaxies as the spherical region enclosed within a radius of $2\, \mathrm{kpc}$. We use this definition of central regions as a proxy of the bulge component of our galaxies, which are selected within the Milky Way stellar mass range at all redshift. We note that this radius of $2\, \mathrm{kpc}$ is consistent with the half-light radii distribution obtained by \citet{Gargiulo2022} for a sample of Milky Way-like galaxies in the TNG50 simulation. On the other hand, we define the disk component as the cylindrical region enclosed within the radial range $2\, \mathrm{kpc} < r < R_{90}$ and vertical distance above and below the disk plane $-h_{90} < z < +h_{90}$.
Similarly, we also define the SFR of the central component as the total SFR of all gas cells enclosed within a sphere of radius $2\, \mathrm{kpc}$, while we define the SFR of the disk as the total SFR of all gas cells enclosed within a cylindrical region of radius $2\, \mathrm{kpc} < r < R_{90}$ and vertical distance above and below the disk plane $-h_{90} < z < +h_{90}$.

\subsection{Calculating the kinematic properties of the galaxies}
\label{sec:kinematic_properties}
The spin-ellipticity (i.e. $\lambda_{\mathrm{R}}$-$\epsilon$) diagram is typically used in observations to separate between fast-rotating and slow-rotating galaxies. The former are typically characterized by a disk-like structure (i.e. lenticular and spiral galaxies), while the latter are typically elliptical galaxies \citep{Emsellem2007,Emsellem2011}. For all the selected galaxies in Sec. \ref{sec:sample_selection}, we use the $\lambda_{\mathrm{R}}$-$\epsilon$ diagram to select our final sample of disk-like galaxies.
As described in \citet{Lagos2016}, we calculate the stellar angular momentum of galaxies, $\lambda_{\mathrm{R}}$, which is defined as:

\begin{equation}
    \lambda_{\mathrm{R}} = \frac{\sum_{{\rm j=1}}^{N} M_{{\rm j}} R_{{\rm j}} V_{\mathrm{rot}} (R_{{\rm j}})}{\sum_{{\rm j}} M_{{\rm j}} R_{{\rm j}} \sqrt{ (V_{\mathrm{rot}} (R_{{\rm j}}))^{2} + (\sigma_{\mathrm{1D}}(R_{{\rm j}}))^{2} }},
\label{eq:lambdaR}
\end{equation}

where $V_{\mathrm{rot}}(R_{j})$ and $\sigma_{\mathrm{1D}}(R_{{\rm j}})$ represent the galaxy rotational velocity and stellar velocity dispersion perpendicular to the mid-plane of the galactic disk calculated within the radius $R_{j}$, respectively. $M_{j}$ represents the stellar mass enclosed within $R_{j}$. 
In Eq. \ref{eq:lambdaR}, the sum runs over all the $N$ radial bins. However, in this work, we consider one single bin that extends from the galaxy center to its stellar half-mass radius (i.e. $R_{j} = 1\, R_{\mathrm{h}}$), which we use as a proxy of the stellar half-light radius. Then, $M_{j}$ represents the stellar mass enclosed within $R_{\mathrm{h}}$, while $V_{\mathrm{rot}}(R_{j})$ represents the galaxy rotational velocity at $R_{\mathrm{h}}$.
The galaxy rotation velocity is then defined as:

\begin{equation}
    V_{\mathrm{rot}} (r) \equiv \frac{|{\bf j}_{*}|}{r},
\label{eq:rotational_velocity}
\end{equation}

where $r = R_{\mathrm{h}}$ and ${\bf j}_{*}$ is the specific angular momentum of the stellar component defined as:

\begin{equation}
    {\bf j}_{*} = \frac{\sum_{{\rm i}} m_{{\rm i}} ( {\bf r}_{{\rm i}} - {\bf r}_{{\rm COM}} ) \times ( {\bf v}_{{\rm i}} - {\bf v}_{{\rm COM}})}{\sum_{{\rm i}} m_{{\rm i}}},
\label{eq:angular_momentum}
\end{equation}

where ${\bf r}_{{\rm i}}$, ${\bf v}_{{\rm i}}$ and ${\bf r}_{{\rm COM}}$, ${\bf v}_{{\rm COM}}$ are the position and velocity vectors of the i-th stellar particle and center of mass of the galaxy, respectively, while $m_{i}$ is the mass of the i-th stellar particle.

The stellar velocity dispersion perpendicular to the mid-plane of the disks is calculated by considering the component of the velocity vector parallel to the total stellar angular momentum vector of the galaxy ${\bf L}_{{\rm *}}$. Then, the stellar velocity dispersion is defined as:

\begin{equation}
    \sigma_{\mathrm{1D},*}(r) = \sqrt{ \frac{\sum_{{\rm i}} m_{{\rm i}} (\Delta v_{{\rm i}} \cos{\theta_{\mathrm{i}}})^{2}}{\sum_{{\rm i}} m_{{\rm i}}} },
\label{eq:velocity_dispersion}
\end{equation}

where $\Delta v_{{\rm i}}$ is the velocity of the i-th stellar particle relative to the centre of mass of the galaxy, while $\cos{\theta_{\mathrm{i}}} = {\Delta \bf v}_{{\rm i}} \cdot {\bf L}_{{\rm *}} / |{\Delta \bf v}_{{\rm i}}| |{\bf L}_{{\rm *}}|$. 

We note that all quantities in Eqs. \ref{eq:lambdaR}, \ref{eq:rotational_velocity}, \ref{eq:angular_momentum} and \ref{eq:velocity_dispersion} are calculated within the inner $R_{\mathrm{h}}$ due to the fact that we are measuring $\lambda_{\mathrm{R}}$ within $R_{\mathrm{h}}$. The total stellar angular momentum of the galaxy is the only quantity calculated using all stellar particles bound to the galaxy. This quantity is only used in Eq. \ref{eq:velocity_dispersion} to define the angle between the stellar velocity and total angular momentum vector of the galaxy to calculate the stellar velocity dispersion perpendicular to the mid-plane of the disk.

For each one of the selected galaxies in Sec. \ref{sec:sample_selection}, we also calculate the ellipticity within the inner $R_{\mathrm{h}}$. To do this, we construct the tensor of inertia from the stellar particles, which, due to its symmetry, is defined as:

\begin{equation}
    \begin{split}
       I_{xx} (r) &= \frac{\sum_{{\rm i}} m_{{\rm i}} (y_{{\rm i}}^{2} + z_{{\rm i}}^{2})}{\sum_{{\rm i}} m_{{\rm i}}} \\
       I_{yy} (r) &= \frac{\sum_{{\rm i}} m_{{\rm i}} (x_{{\rm i}}^{2} + z_{{\rm i}}^{2})}{\sum_{{\rm i}} m_{{\rm i}}} \\
       I_{zz} (r) &= \frac{\sum_{{\rm i}} m_{{\rm i}} (x_{{\rm i}}^{2} + y_{{\rm i}}^{2})}{\sum_{{\rm i}} m_{{\rm i}}} \\
       I_{xy} (r) &= I_{yx} (r) = - \frac{\sum_{{\rm i}} m_{{\rm i}} (x_{{\rm i}}y_{{\rm i}})}{\sum_{{\rm i}} m_{{\rm i}}} \\
       I_{xz} (r) &= I_{zx} (r) = - \frac{\sum_{{\rm i}} m_{{\rm i}} (x_{{\rm i}}z_{{\rm i}})}{\sum_{{\rm i}} m_{{\rm i}}} \\
       I_{yz} (r) &= I_{zy} (r) = - \frac{\sum_{{\rm i}} m_{{\rm i}} (y_{{\rm i}}z_{{\rm i}})}{\sum_{{\rm i}} m_{{\rm i}}}. 
    \end{split}
\label{eq:inertia_tensor}
\end{equation}

Then, we diagonalize the tensor of inertia to derive the corresponding eigenvalues, $M_{1}$, $M_{2}$ and $M_{3}$, which are sorted in ascending order, such as $M_{1} < M_{2} < M_{3}$, and used to define the galaxy shortest-to-longest axis ratio $c/a = M_{1}/M_{3}$ \citep{Genel2015}.
Finally, the ellipticity of the galaxy is defined as:

\begin{equation}
    \epsilon = 1 - \frac{c}{a},
\end{equation}

\subsection{Defining the sample of disk-like galaxies}
\label{sec:selected_disk_galaxies}
\begin{figure}
    \centering
    \includegraphics[width=0.5\textwidth]{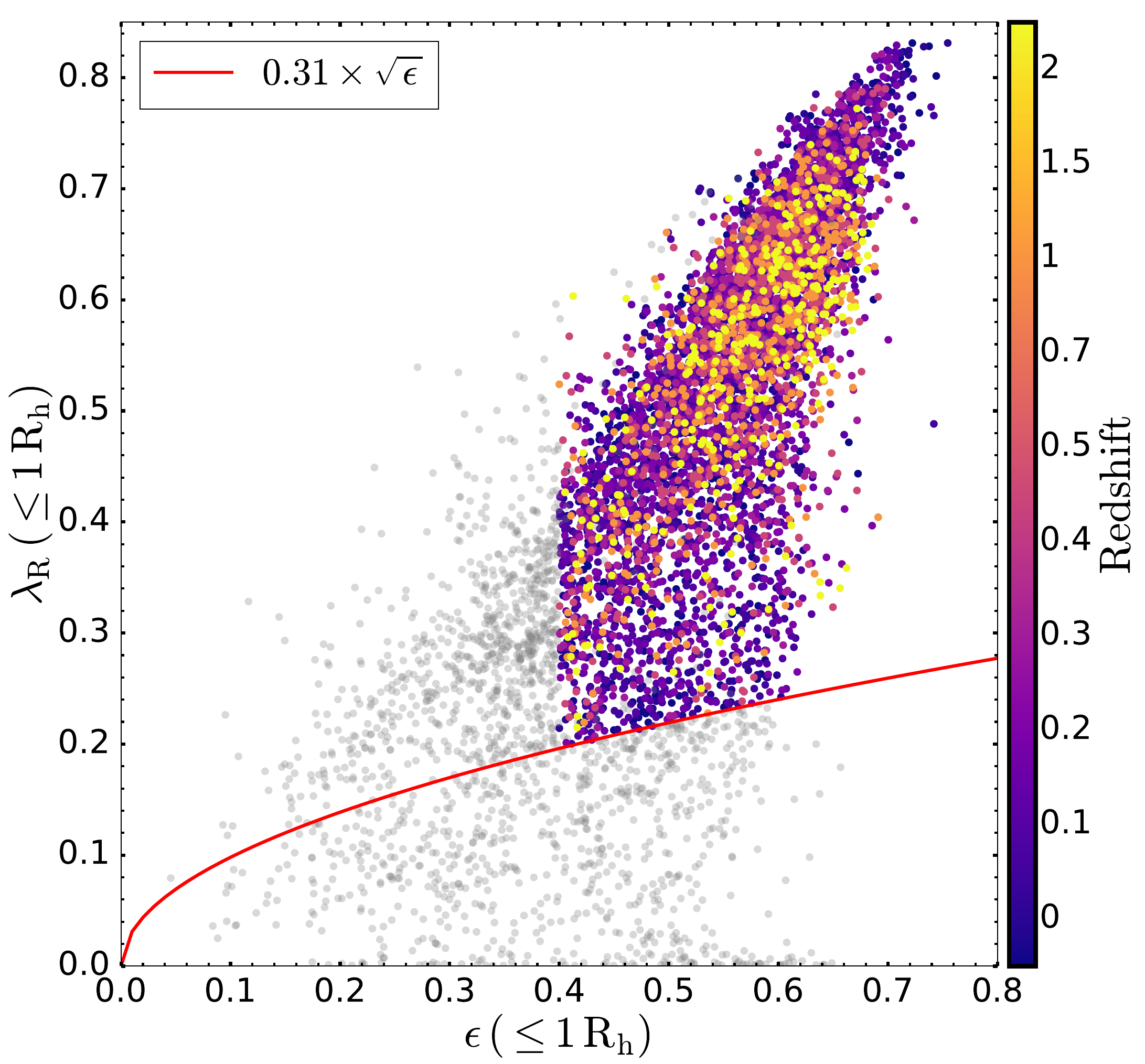}
    \includegraphics[width=0.45\textwidth]{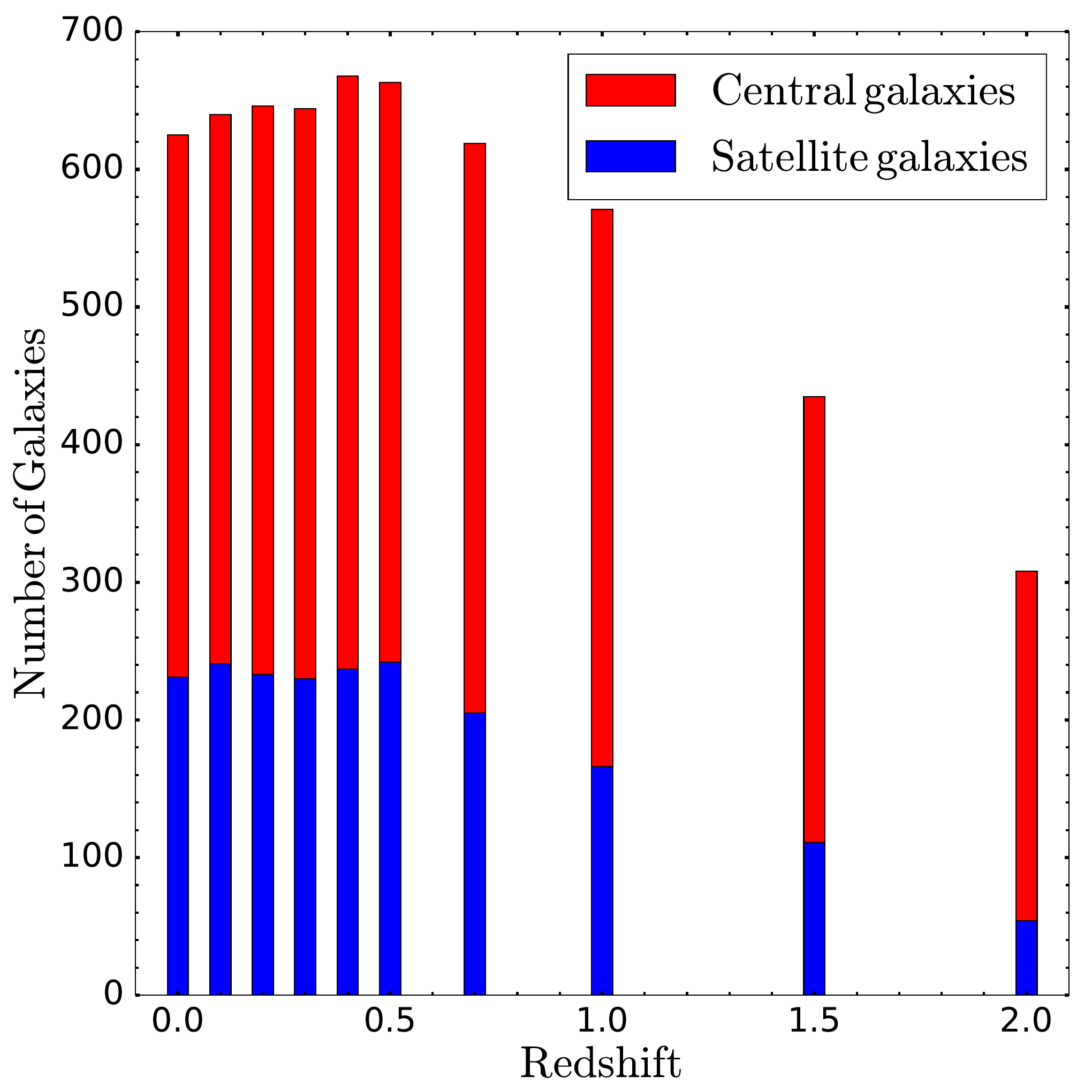}
    \caption{{\it Top panel:} Spin-ellipticity (i.e. $\lambda_{\mathrm{R}}$-$\epsilon$) diagram of the galaxies selected from the TNG50 simulation between $0 < z < 2$ (gray points). The colored points show our final sample of disk-like galaxies, using the selection criteria described in Sec. \ref{sec:selected_disk_galaxies}. The red solid line represents the threshold typically used in observations to separate between rotation-dominated and dispersion-supported galaxies within the inner stellar half-light radius at $z=0$ \citep{Emsellem2011}. {\it Bottom panel:} The total number of central and satellite galaxies of our final sample of disk-like galaxies at each redshift, selected as described in Sec. \ref{sec:selected_disk_galaxies}.}
    \label{fig:spin-ellipticity_diagram}
\end{figure}

In the top panel of Fig. \ref{fig:spin-ellipticity_diagram}, we show the spin-ellipticity (i.e. $\lambda_{\mathrm{R}}$-$\epsilon$) diagram for all the selected galaxies in Sec. \ref{sec:sample_selection}. The red solid line shows the threshold $\lambda_{\mathrm{R}} = 0.31\times\sqrt{\epsilon}$ typically used in observations to separate between rotation-dominated and dispersion-supported galaxies within the inner stellar half-light radius at $z=0$ \citep{Emsellem2011}, which we assume to be valid up to $z=2$. 
We select our final sample of disk-like galaxies by including all rotation-dominated (i.e. $\lambda_{\mathrm{R}} > 0.31\times\sqrt{\epsilon}$) and flattened (i.e. $\epsilon > 0.4$) galaxies with $R_{90} > 3\, \mathrm{kpc}$ at each redshift between $0 < z < 2$ (colored points in Fig. \ref{fig:spin-ellipticity_diagram}). The $\epsilon > 0.4$ selection criteria is an additional constraint for selecting flattened galaxies with disk morphology, consistent with spiral galaxies, similarly to \citet{Gargiulo2022}. After visual inspection, we also remove compact galaxies with $R_{90} < 3\, \mathrm{kpc}$ that did not show evidence of a disk-like structure from the edge-on projection. These galaxies represent a minority (i.e. $17$ at $z=2$ and only a few at $z=0$). In the bottom panel of Fig. \ref{fig:spin-ellipticity_diagram}, we show the number of central and satellite galaxies of our final selected sample of disk-like galaxies at each redshift. We see that the fraction of satellite galaxies increases from $z=2$ up to $z=0.5$ and, then, it remains overall constant up to $z=0$. 

\begin{figure}
    \centering
    \includegraphics[width=0.45\textwidth]{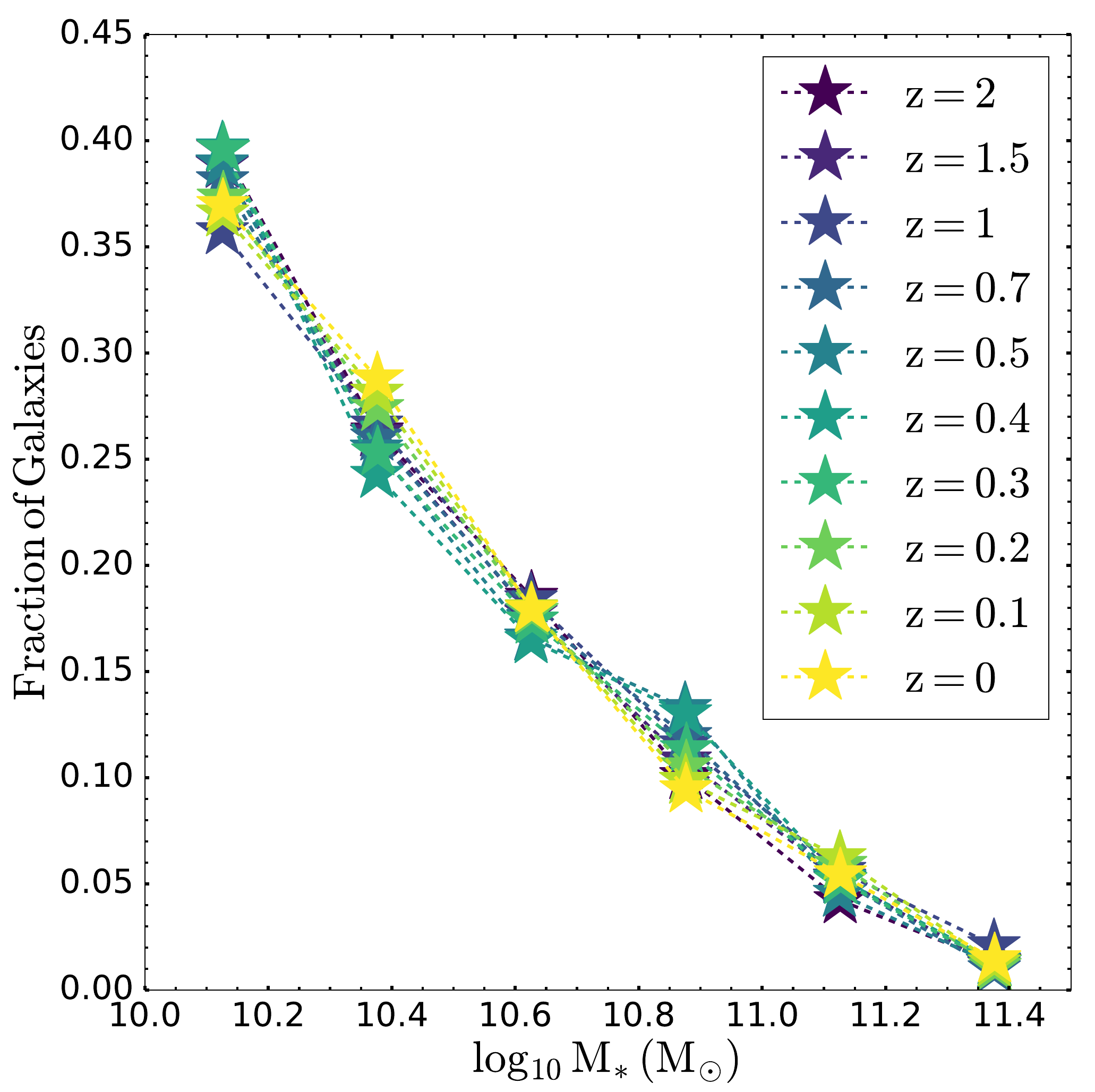}
    \includegraphics[width=0.45\textwidth]{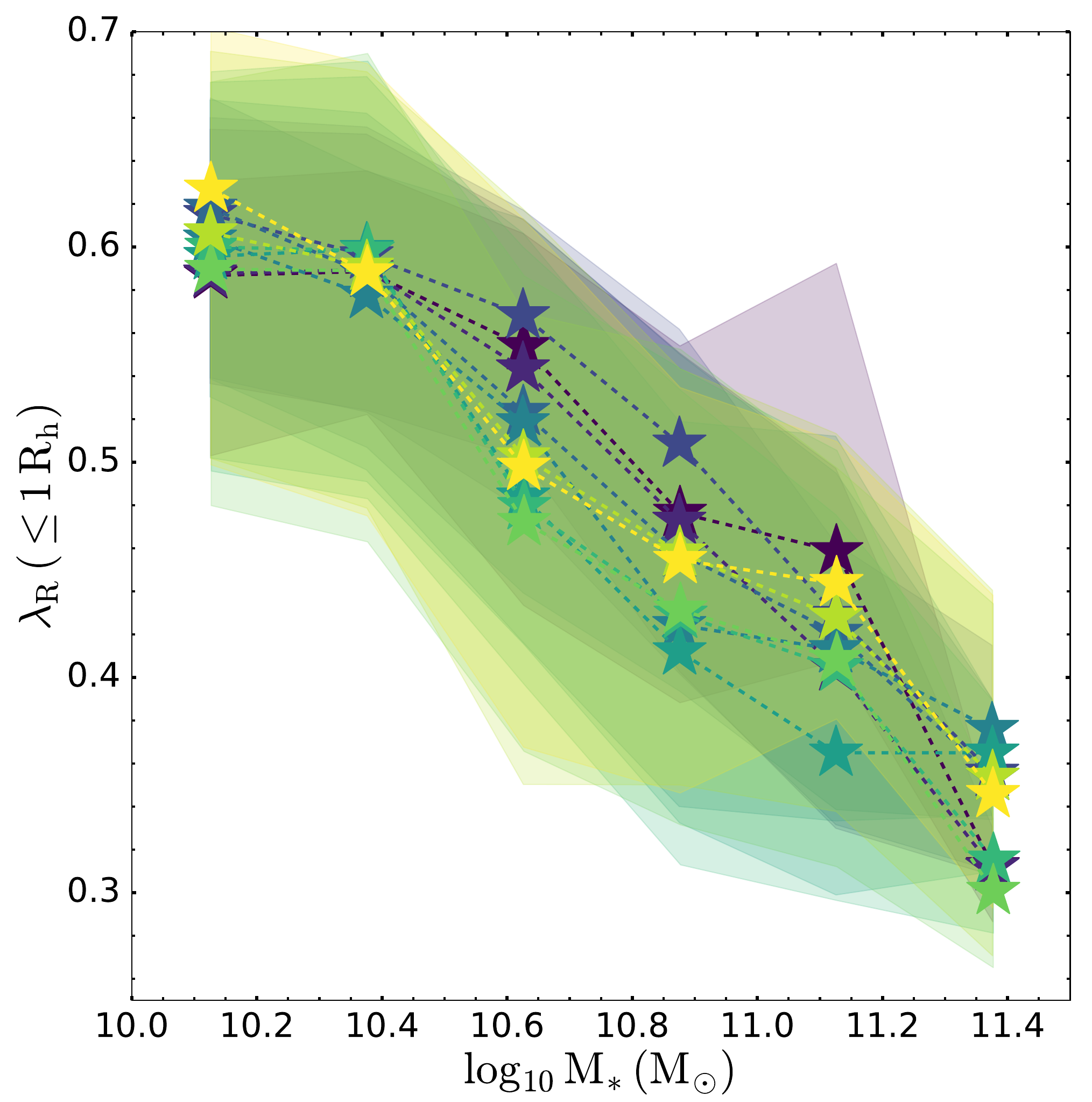}
    \caption{Stellar mass distribution (top panel) and median value of $\lambda_{\mathrm{R}}$ as a function of stellar mass (bottom panel) for our selected sample of disk-like galaxies at each redshift between $0 < z < 2$. Shaded areas are defined by the $25$th-$75$th interquartile range of the data in each bin.}
    \label{fig:selected_disk_galaxies}
\end{figure}

In Fig. \ref{fig:selected_disk_galaxies}, we show the stellar mass distribution and the median value of $\lambda_{\mathrm{R}}$ as a function of stellar mass for our selected sample of disk-like galaxies at each redshift between $0 < z < 2$. Overall, we see that the fraction of disk-like galaxies, as well as their degree of rotation, steeply decreases towards high stellar masses at all redshift. This is consistent with the observations that massive galaxies tend to be slow-rotating spheroidals \citep{Emsellem2011}.

\begin{figure*}
    \centering
    \includegraphics[width=0.45\textwidth]{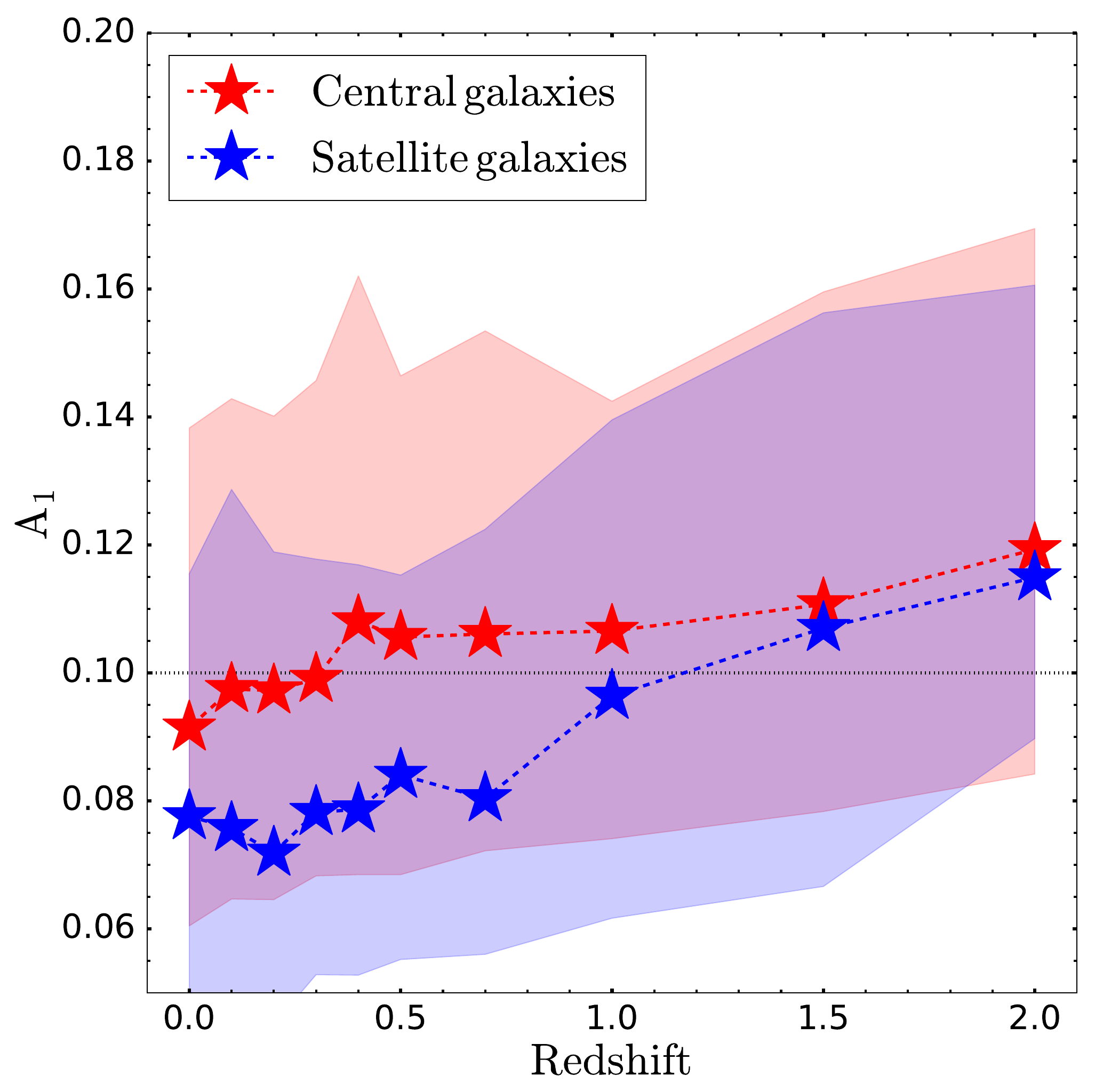}
    \includegraphics[width=0.45\textwidth]{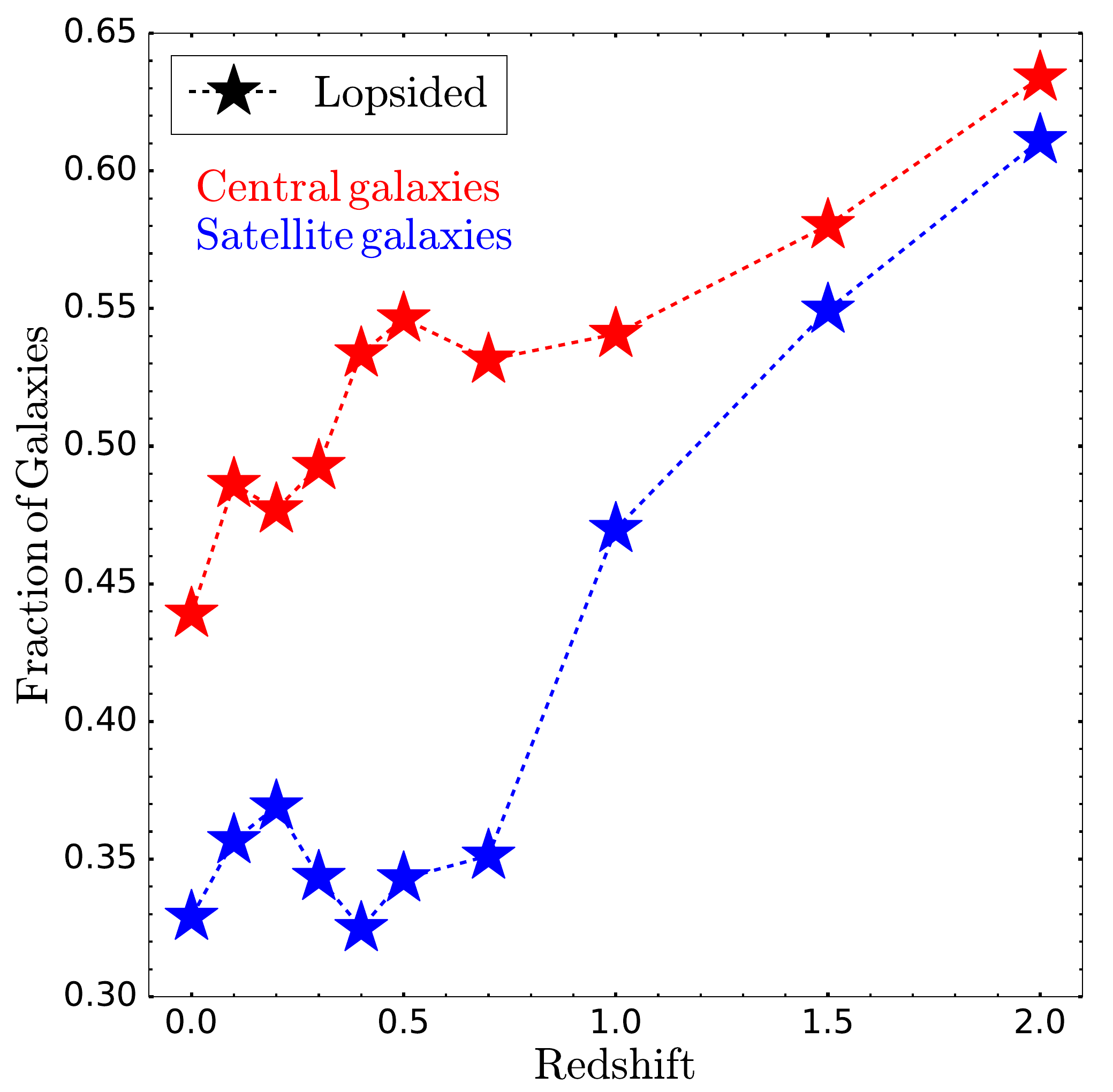}
    \caption{{\it Left panel:} Median lopsided amplitude as a function of redshift for our selected sample of disk-like galaxies at each analyzed redshift between $0 < z < 2$, divided into centrals and satellites. Shaded areas are defined by the $25$th-$75$th interquartile range of the data at each bin. The horizontal dotted black line indicates the threshold $\mathrm{A}_{1}=0.1$ used to classify galaxies into lopsided ($\mathrm{A}_{1}>0.1$) and symmetric ($\mathrm{A}_{1}<0.1$) at $z=0$ \citep{Varela-Lavin2023,Dolfi2023}. {\it Right panel:} Fraction of lopsided galaxies as a function of redshift for centrals and satellites.}
    \label{fig:lopsidedness_evolution}
\end{figure*}

\section{Measuring lopsidedness}
\label{sec:lopsidedness}
Similarly to \citet{Varela-Lavin2023} and \citet{Dolfi2023}, we calculate the radial lopsidedness profile of our selected sample of disk-like galaxies at each redshift between $0 < z < 2$ by performing an azimuthal Fourier decomposition of the galaxy stellar mass in the face-on projection. We measure the amplitude of the first Fourier mode, which we define here as lopsided amplitude, in equally-spaced concentric radial annuli of width $0.1\, \mathrm{kpc}$ and height $|h_{z}| = 2\, h_{90}$ above and below the disk plane. Finally, we quantify the global lopsidedness, $\mathrm{A}_{1}$, of our galaxies by calculating the average lopsided amplitude in the radial range $R_{\mathrm{h}} < r < 1.4\, R_{90}$.

We note that the radial range $R_{\mathrm{h}} < r < 1.4\, R_{90}$ differs from the one used in \citet{Varela-Lavin2023} and \citet{Dolfi2023} to quantify the global lopsidedness of galaxies at $z=0$, i.e. $0.5\, R_{\mathrm{opt}} < r < 1.1\, R_{\mathrm{opt}}$ with $R_{\mathrm{opt}}$ being the optical radius defined as the radius where the $V$-band surface brightness drops down to $\mu_{V}=26.5\, \mathrm{mag}/\mathrm{arcsec^{2}}$ at $z=0$. 
In this work, we find that the outer radius $1.4\, R_{90}$ is a good proxy of the boundary of the stellar disks of galaxies at all redshift up to $z=2$. On the other hand, we find that, if we use the same $z=0$ definition of $R_{\mathrm{opt}}$ for all redshift, then $R_{\mathrm{opt}}$ is more extended than the boundary of galactic disks at high-redshift (i.e. $z\gtrsim0.5$).
In any case, we have studied the effect of using different radial ranges to quantify the global lopsidedness of galaxies between $0 < z < 2$. Overall, we find that, while the fraction of lopsided and symmetric galaxies varies depending on the radial range used (see also discussion in \citealt{Dolfi2023} at $z=0$), the general trends and results are not significantly affected.

\section{Results}
\label{sec:results}
Throughout the discussion of the results, we remind the reader that, in this work, we are not following the evolution of a certain sample of galaxies, but rather we are investigating the behavior of a defined population of disk-like galaxies at each redshift. 

\subsection{Lopsidedness as a function of redshift}
\label{sec:lopsidedness_evolution}
In the left panel of Fig. \ref{fig:lopsidedness_evolution}, we show the median value of the lopsided amplitude as a function of redshift for our selected sample of disk-like galaxies between $0 < z < 2$, divided into centrals and satellites. Overall, we see a decrease of the lopsided amplitude from $z=2$ to $z=0$ for both central and satellite galaxies, but this decrease is less significant for central galaxies. Additionally, we see that, while central and satellite galaxies tend to be characterized by a similar lopsided amplitude at high-redshift (i.e. $z>1$), central galaxies have larger lopsided amplitude than satellite galaxies at low-redshift (i.e. $z\lesssim1$).

In the right panel of Fig. \ref{fig:lopsidedness_evolution}, we show the fraction of lopsided galaxies as a function of redshift for the central and satellite galaxies, respectively. Here, we use the threshold $\mathrm{A}_{1}=0.1$ to classify galaxies into lopsided ($\mathrm{A}_{1}>0.1$) and symmetric ($\mathrm{A}_{1}<0.1$) at all redshift, as previously done in \citet{Varela-Lavin2023} and \citet{Dolfi2023} at $z=0$. At high-redshift (i.e. $z>1$), we find similar large fractions of lopsided galaxies (i.e. $\gtrsim55\%$) for both centrals and satellites. This is overall consistent with the large fraction of observed lopsided galaxies (i.e. $64\%$) found by \citet{LeBail2023} at high-redshift (i.e. $1.5 \lesssim z \lesssim 3$). For $z\lesssim1$, we find that the fraction of lopsided satellite galaxies steeply decreases to $\sim30$-$35\%$ at $z=0$. On the other hand, the fraction of lopsided central galaxies remains roughly constant at $\sim55\%$ up to $z\sim0.5$, beyond which it drops to $\sim45\%$ at $z=0$.   
We note that, at $z\sim0$, we find a slightly larger number of lopsided galaxies (i.e. $\sim40\%$) with respect to observations in the local Universe (i.e. $\sim30\%$; e.g. \citealt{Zaritsky1997}). 
The reason for this discrepancy between observations and simulations can be likely attributed to the radial range used to measure the global lopsidedness of galaxies. While \citet{Zaritsky1997} measured lopsidedness by taking the average amplitude between $1.5$-$2.5$ disk scalelengths (i.e. $\sim1$-$2\, R_{\mathrm{h}}$; \citealt{Lokas2022}), we measure lopsidedness reaching out to larger galactocentric radii (i.e. up to $1.4\, R_{90}$). For this reason, we are likely identifying a larger fraction of lopsided galaxies compared to observations, due to the fact that the strength of the lopsided perturbation typically increases towards large galactocentric radii \citep{Varela-Lavin2023}. Indeed, the estimated fraction of lopsided galaxies is sensitive to the radial range used to measure the global lopsidedness (see also the discussion in \citealt{Dolfi2023}).

Overall, the results shown in Fig. \ref{fig:lopsidedness_evolution} suggest that galaxies tend to become more symmetric towards low-redshift, and that mechanisms generating lopsidedness affect similarly central and satellite galaxies at high-redshift (i.e. $z>1$), while they are not as efficient for satellite galaxies close to the present-day (i.e. $z\lesssim1$). Environmental interactions could be playing a more dominant role in the origin of lopsidedness at high-redshift, as suggested by the high fraction of lopsided satellite galaxies, while other mechanisms could be primarily triggering lopsidedness at low-redshift. In general, the decrease in the fraction of lopsided galaxies towards low-redshift is consistent with the observations from the morphology-density relation \hbox{\citep{Dressler1980,Dressler1997}}. In galaxy groups and clusters, the fraction of early-type disk galaxies (i.e. S0-like) generally increases with increasing environmental density, as well as from high-redshift (i.e. $z\sim0.5$) to low-redshift, as a result of environmental effects driving morphological transformations. As previously shown in \citet{Varela-Lavin2023} and \citet{Dolfi2023} at $z=0$, early-type disk galaxies have typically lower lopsided amplitude than late-type ones, due to the fact that they are more gravitationally cohesive and, thus, less affected by external perturbations \citep{Varela-Lavin2023}. For this reason, we find that satellite galaxies are more symmetric than central galaxies at low-redshift, due to the fact that they are experiencing more rapid morphological transformations to early-type disks.

\begin{figure}
    \centering
    \includegraphics[width=0.45\textwidth]{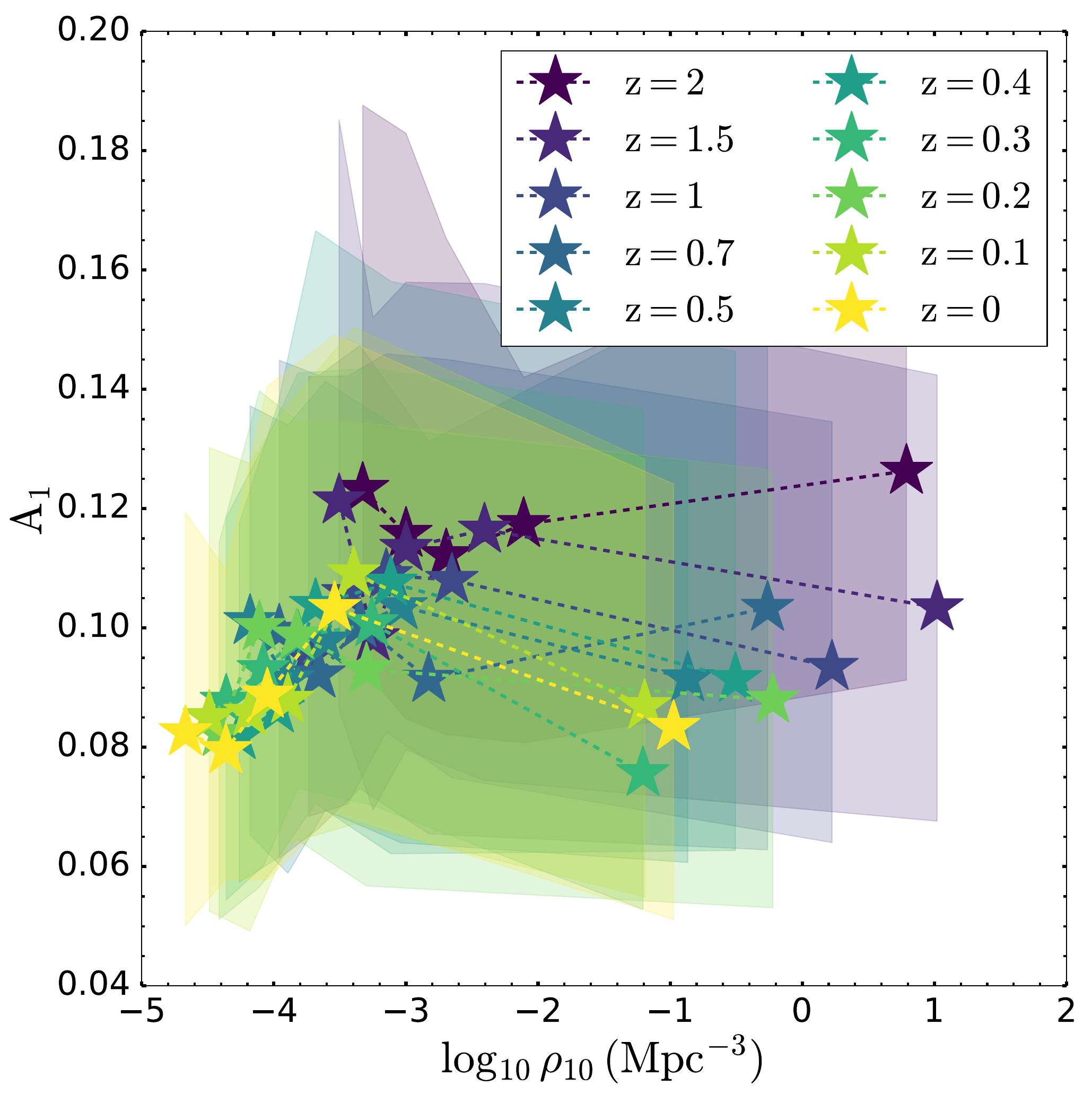}
    \caption{The median lopsided amplitude as a function of the local density of the environment between $0 < z < 2$ for our selected sample of disk-like galaxies. The local density of the environment is calculated as described in Sec. \ref{sec:lopsidedness_environment}, considering the ten nearest neighbors with total mass $\mathrm{M}_{\mathrm{tot}} > 10^{9}\, \mathrm{M_{\odot}}$. Shaded areas are defined by the $25$th-$75$th interquartile range of the data in each bin.}
    \label{fig:lopsidedness_local_environment}
\end{figure}

\begin{figure*}
    \centering
    \includegraphics[width=0.33\textwidth]{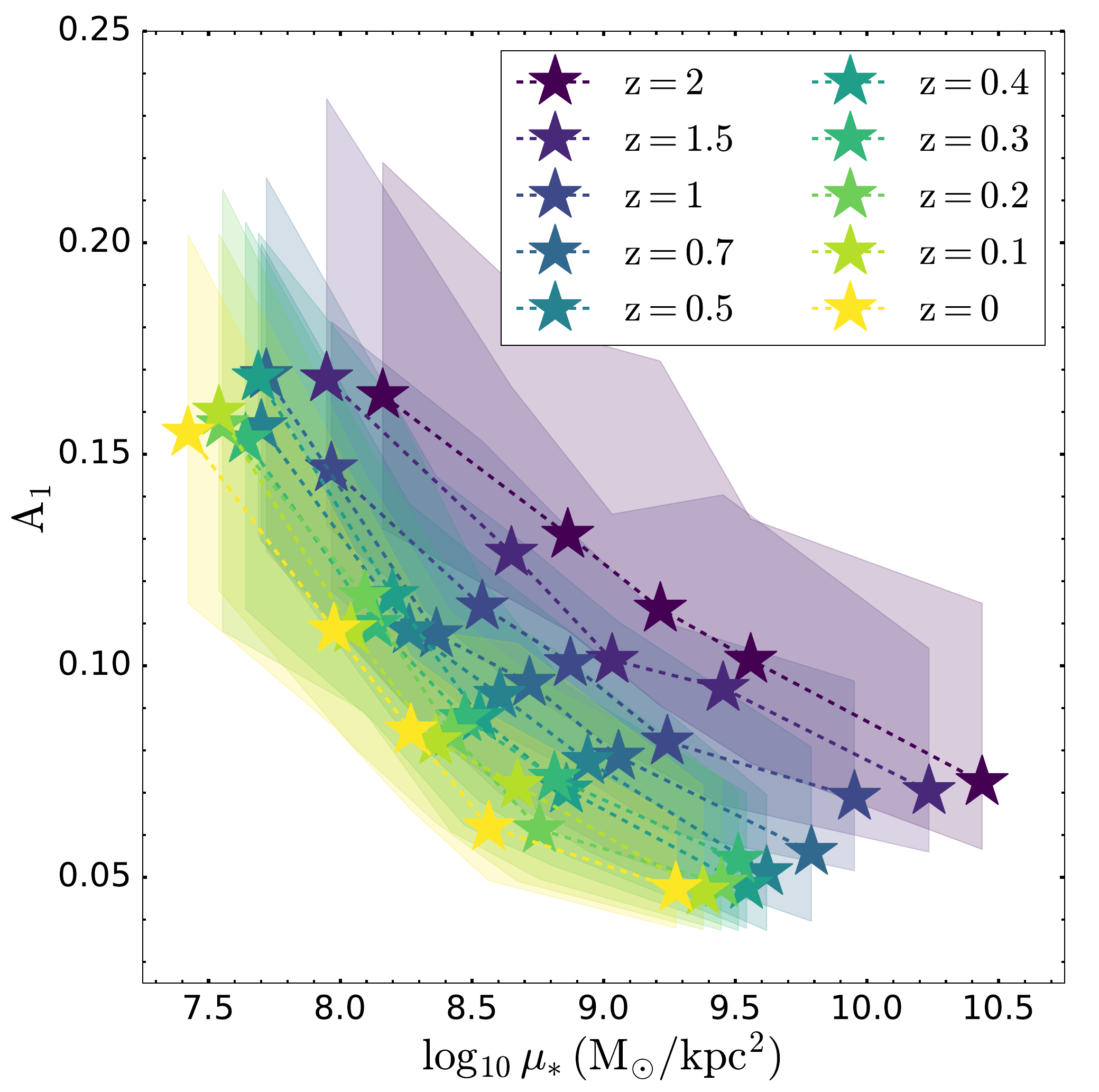}
    \includegraphics[width=0.33\textwidth]{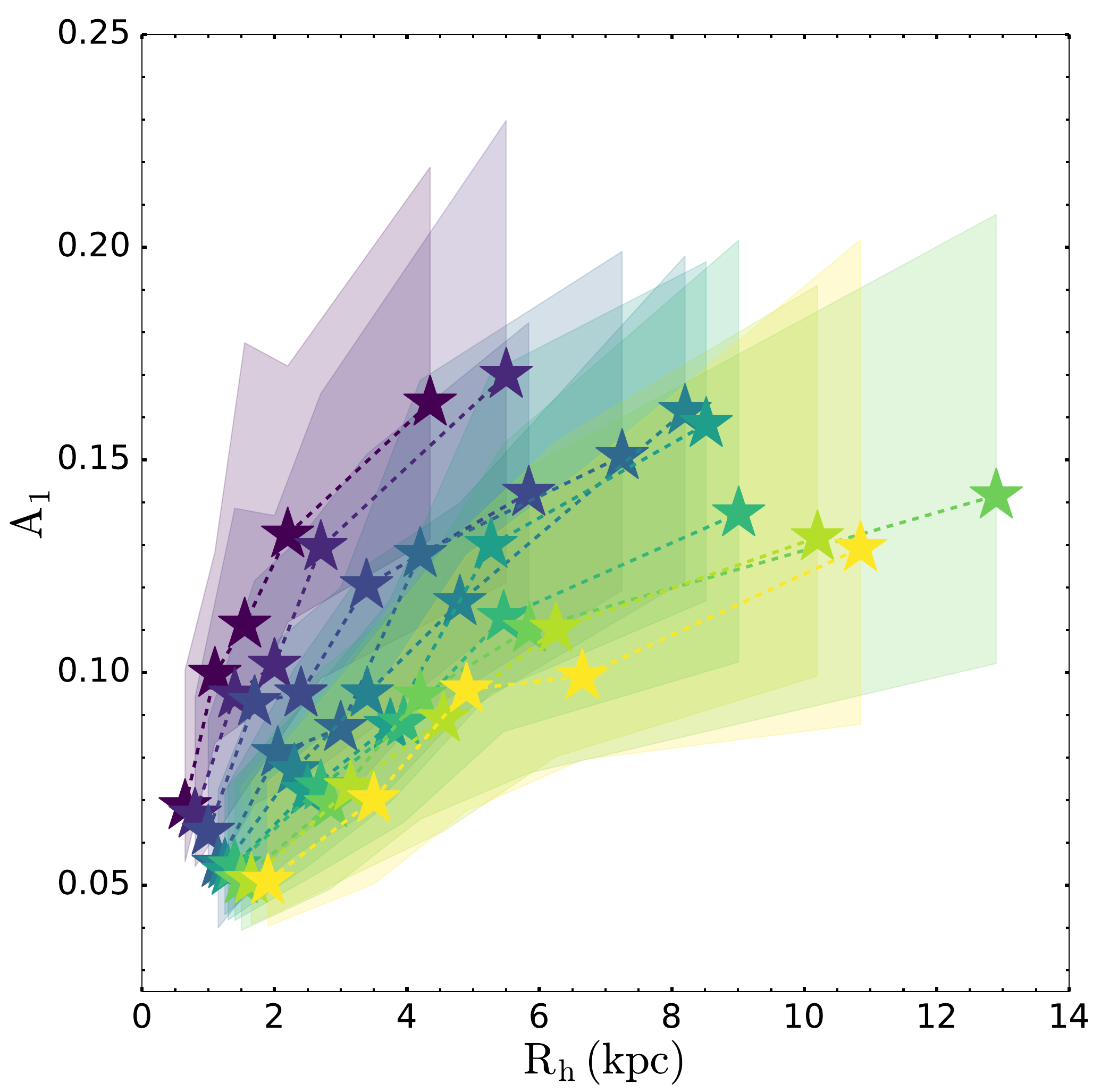}
    \includegraphics[width=0.33\textwidth]{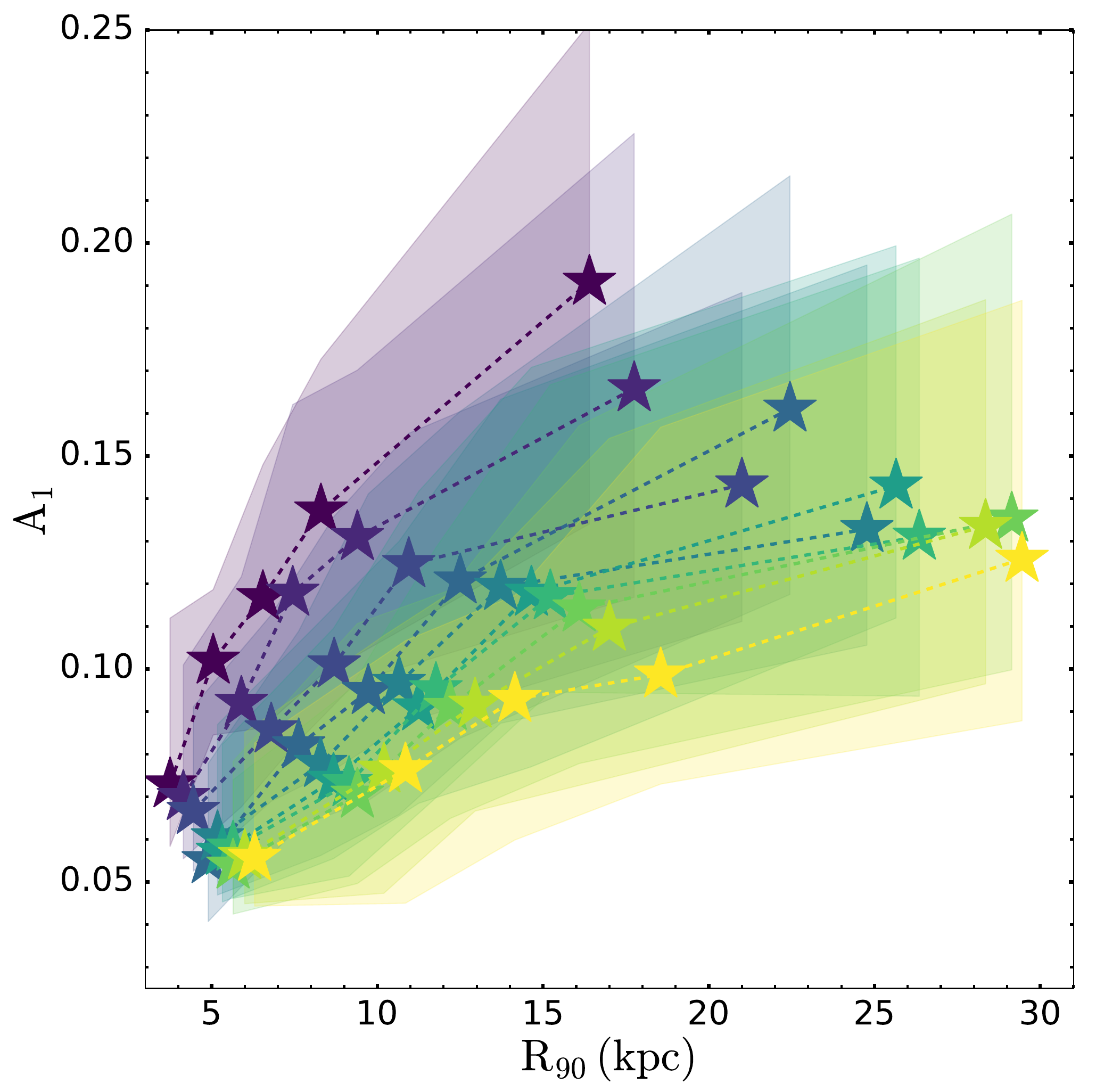}
    \includegraphics[width=0.33\textwidth]{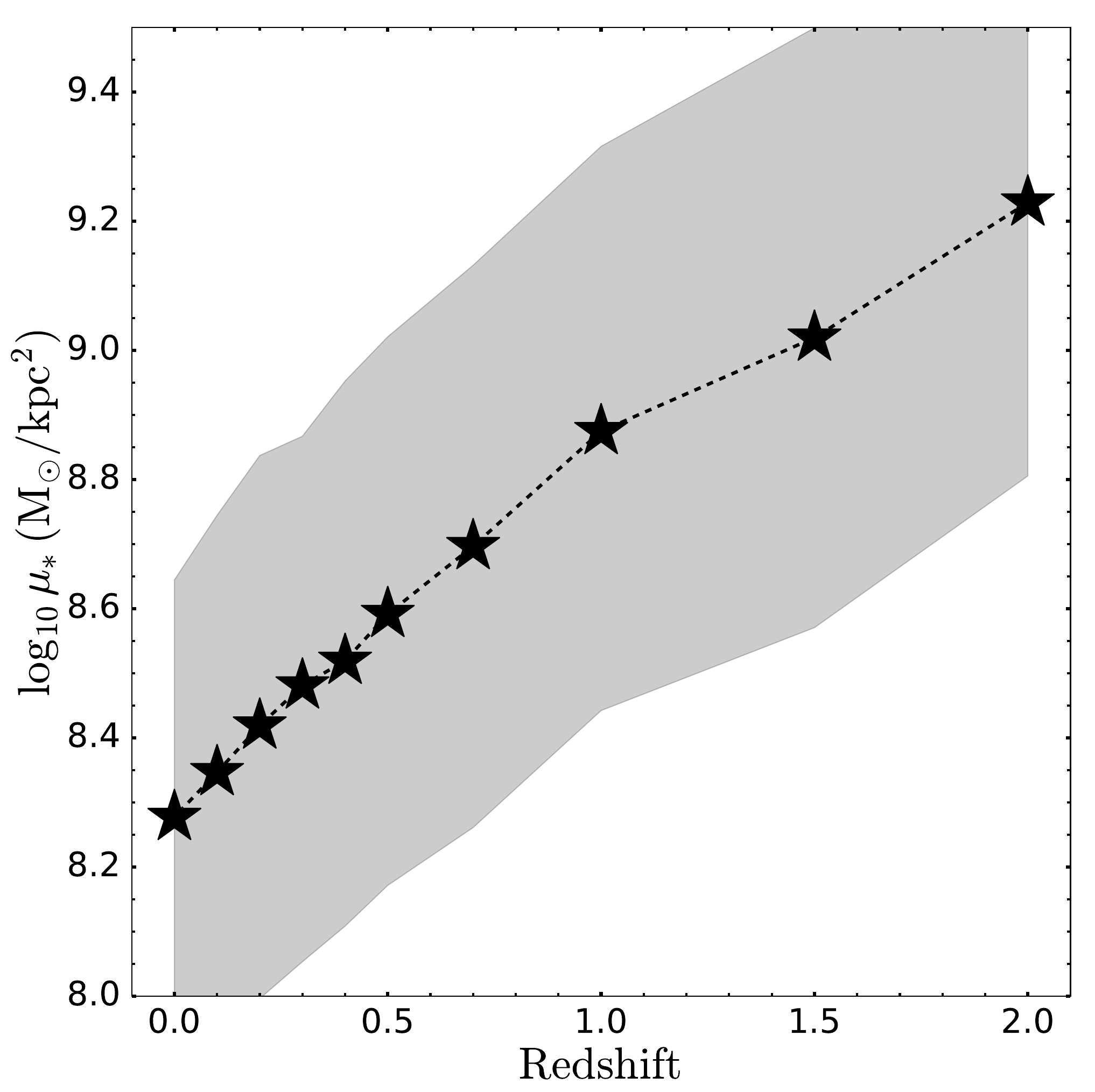}
    \includegraphics[width=0.32\textwidth]{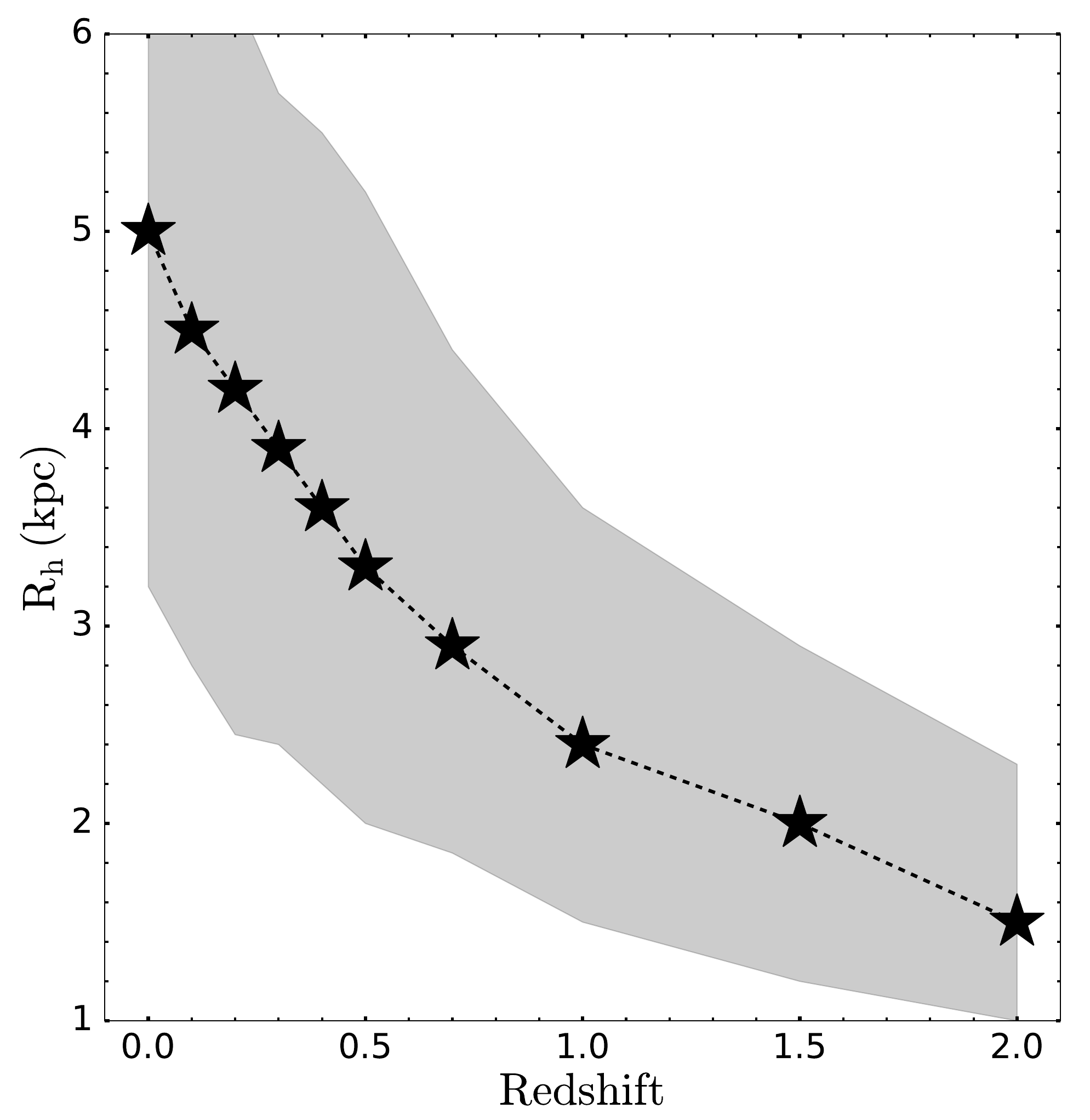}
    \includegraphics[width=0.325\textwidth]{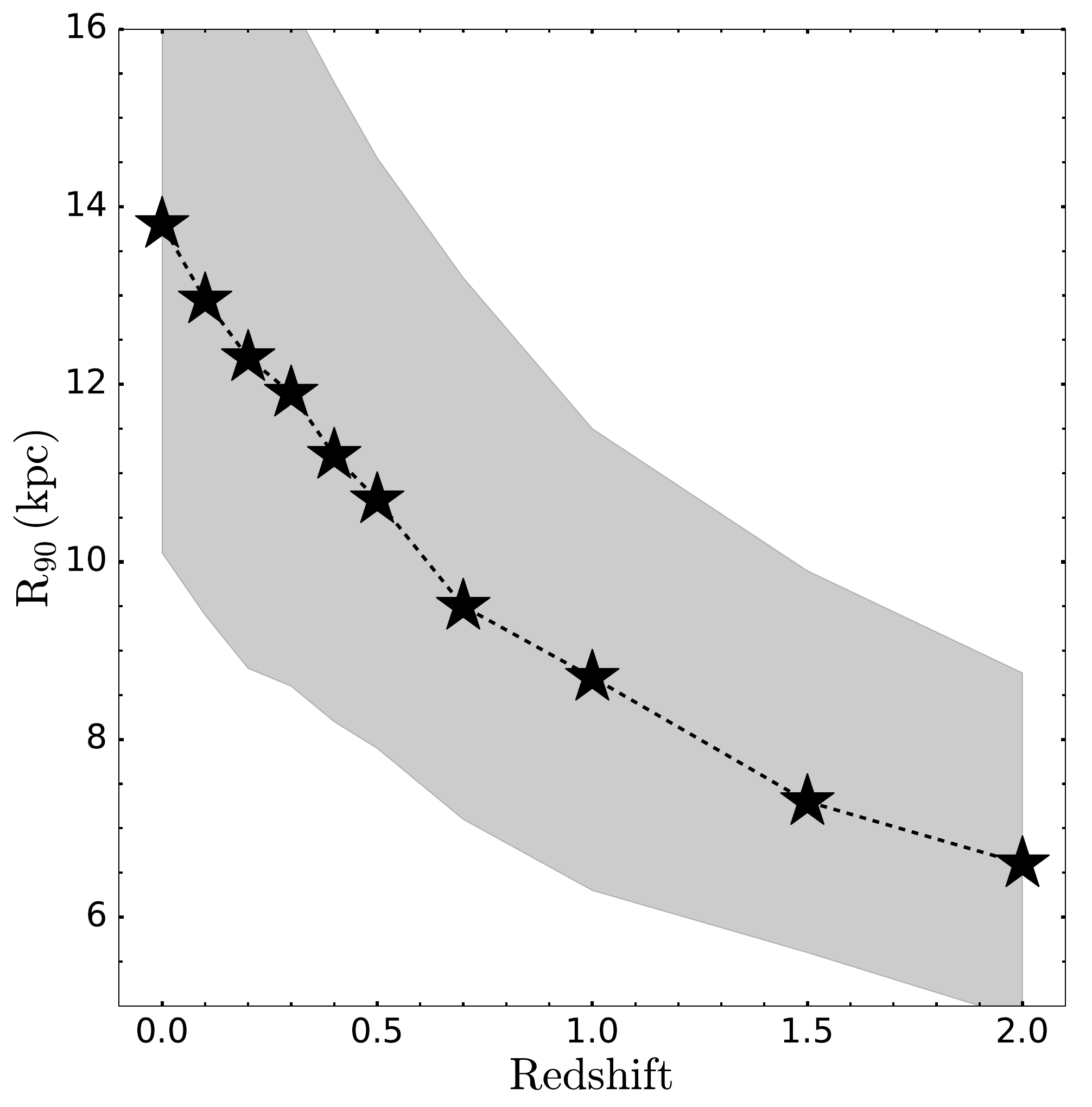}
    \caption{{\it Top:} The median lopsided amplitude as a function of the central stellar mass density ($\mu_{*}$), stellar half-mass radius ($R_{\mathrm{h}}$) and disk size ($R_{90}$) for our selected sample of disk-like galaxies between $0 < z < 2$. {\it Bottom:} The median $\mu_{*}$, $R_{\mathrm{h}}$ and $R_{90}$ as a function of redshift for our selected sample of disk-like galaxies between $0 < z < 2$. Shaded areas are defined by the $25$th-$75$th interquartile range of the data in each bin.}
    \label{fig:lopsidedness_galaxy_properties}
\end{figure*}

\subsection{Correlation between lopsidedness and local environment}
\label{sec:lopsidedness_environment}
We now study the correlation between the lopsided amplitude and the local density of the environment as a function of redshift for our selected sample of disk-like galaxies between $0 < z < 2$. 
We define the local density of the environment around each selected galaxy as:

\begin{equation}
    \rho_{\mathrm{10}} = \mathrm{10}/(\frac{4}{3}\pi R_{\mathrm{10}}^{3}),
\label{eq:density_environment}
\end{equation}

where $R_{\mathrm{10}}$ is the radius of the sphere, centered on each galaxy, enclosing the $\mathrm{N}=10$ nearest neighbors. We only consider nearest neighbors with total mass $\mathrm{M}_{\mathrm{tot}} > 10^{9}\, \mathrm{M_{\odot}}$, where $\mathrm{M}_{\mathrm{tot}}$ is computed considering all particles (i.e. dark matter$+$baryons) that are gravitationally bound to the galaxy. We note here that, in \citet{Dolfi2023}, we tested the use of different environmental metrics, as well as mass thresholds, to study the correlation between the lopsided amplitude and local environment, finding that the results did not vary significantly when using different metrics.

In Fig. \ref{fig:lopsidedness_local_environment}, we show that, independently of the redshift considered, the median lopsided amplitude does not strongly depend on the local environmental density of the galaxies. We find that the results are similar when we consider central and satellite galaxies, separately. 
Overall, this suggests that no particular environment plays a more important role than any other in the origin of lopsidedness, which is generally consistent with the mild or lack of correlation between lopsidedness and local environment observed at $z=0$ both in simulations \citep{Dolfi2023} and observations \citep{Wilcots2010}, as well as in high-redshift observations \citep{LeBail2023}.

\subsection{Correlation between lopsidedness and internal galaxy properties}
\label{sec:lopsidedness_galaxy_properties}
Previous present-day studies of lopsidedness have found that this perturbation strongly correlates with the central stellar mass density and disk size of galaxies \citep{Reichard2008,Varela-Lavin2023}. 

In top panels of Fig. \ref{fig:lopsidedness_galaxy_properties}, we study the correlation between lopsidedness and the internal properties of galaxies as a function of redshift for our selected sample of disk-like galaxies between $0 < z < 2$.
From the left to the right panel of Fig. \ref{fig:lopsidedness_galaxy_properties}, we show the median lopsided amplitude as a function of the central stellar mass density (i.e. $\mu_{*}$\footnote{$\mu_{*}=\mathrm{M_{\mathrm{h}}}/\pi R_{\mathrm{h}}^{2}$, where $\mathrm{M_{\mathrm{h}}}$ is the stellar mass contained within the stellar half-mass radius $R_{\mathrm{h}}$}), stellar half-mass radius ($R_{\mathrm{h}}$) and disk size ($R_{90}$), respectively.  
We find a strong correlation between the median lopsided amplitude and the different internal properties of galaxies at all redshift considered. Galaxies characterized by, on average, lower central stellar mass density, larger half-mass radii and larger $R_{90}$ tend to be more lopsided than galaxies with higher central stellar mass density, smaller half-mass radii and smaller $R_{90}$. However, we also see two additional trends. First of all, the median lopsided amplitude tends to increase from low- to high-redshift at a fixed $\mu_{*}$, $R_{\mathrm{h}}$ and $R_{\mathrm{90}}$. Furthermore, the median lopsided amplitude increases more rapidly with increasing $R_{\mathrm{h}}$ and $R_{90}$ at high- than low-redshift (i.e. steeper $\mathrm{A_{1}}$ slope for $R_{\mathrm{h}}$ and $R_{90}$ at high-redshift). This means that, at high-redshift, lopsidedness is more sensitive to variations in the galaxy's radii.
Overall, these results suggest that the development of the lopsided perturbation is strongly connected to the internal properties of the galaxies at all redshift, consistent with the results from previous works at $z=0$ \citep{Reichard2008,Varela-Lavin2023,Dolfi2023}.

In the bottom panels of Fig. \ref{fig:lopsidedness_galaxy_properties}, we study the evolution of the median $\mu_{*}$, $R_{\mathrm{h}}$ and $R_{90}$ as a function of redshift for our selected sample of disk-like galaxies between $0 < z < 2$. We see that, at high-redshift (e.g. $z>1$), galaxies have larger central stellar mass density and smaller sizes than at lower redshift. This means that, if mechanisms triggering lopsidedness are the same as a function of redshift, low-redshift galaxies should be more susceptible at developing strong lopsidedness than high-redshift ones, due to the observed strong correlation between lopsidedness and galaxy internal properties. The fact that the median lopsided lopsided amplitude, as well as fraction of lopsided galaxies, decrease from high- to low-redshift (see Fig. \ref{fig:lopsidedness_evolution}, thus suggests that mechanisms triggering lopsidedness are different or more efficient at high- than low-redshift.

We also study the correlation between lopsidedness and the galaxy stellar mass. At all redshift considered, we only find a mild decrease of the median lopsided amplitude with increasing stellar mass. This is consistent with the decrease of $\mathrm{A}_{1}$ towards large stellar masses seen in observations at $z=0$ \citep{Reichard2008}, as previously shown in \citet{Dolfi2023}.
We note that we find similar results when we consider central and satellite galaxies, separately.

\begin{figure*}
    \centering
    \includegraphics[width=0.45\textwidth]{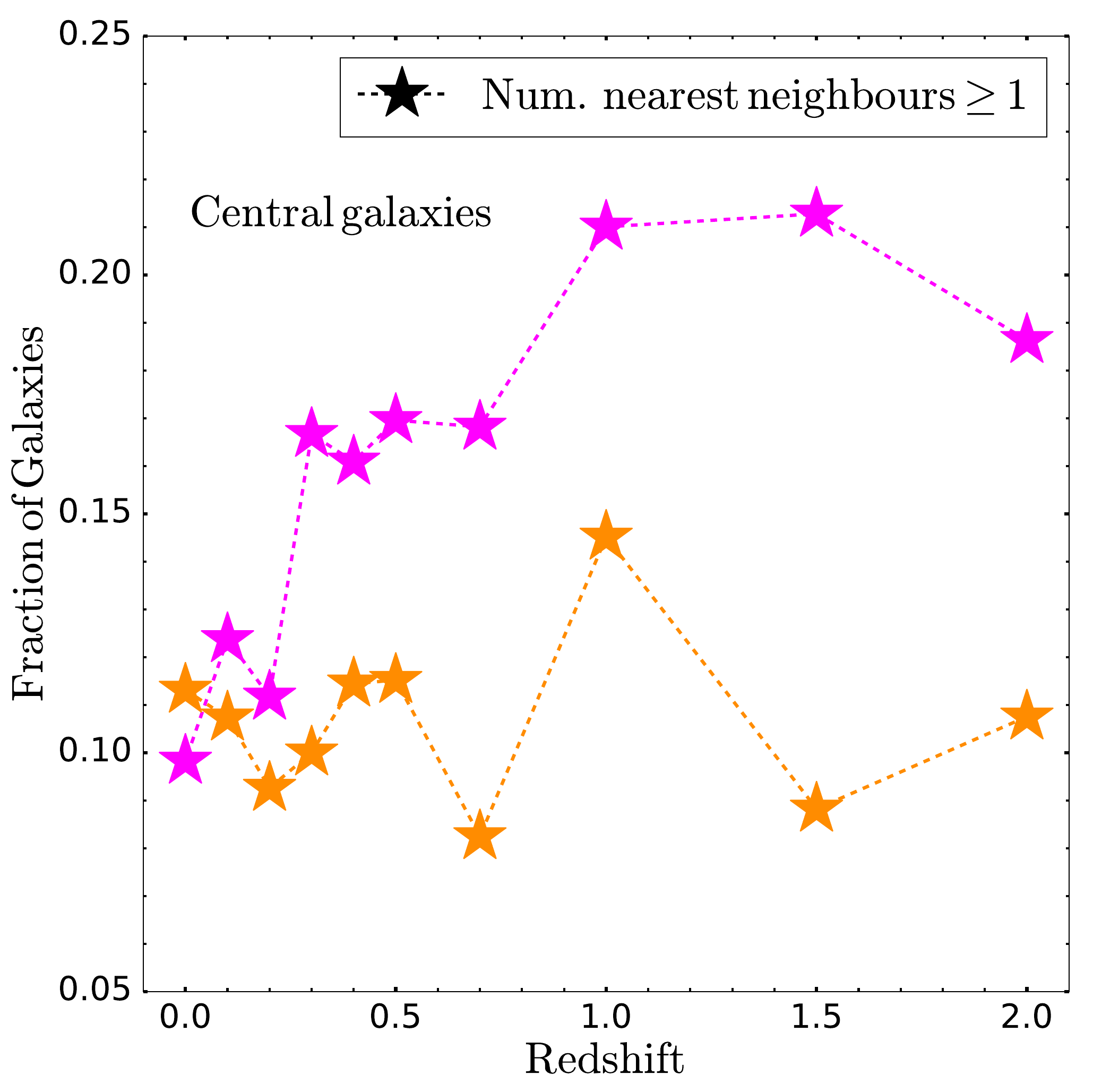}
    \includegraphics[width=0.45\textwidth]{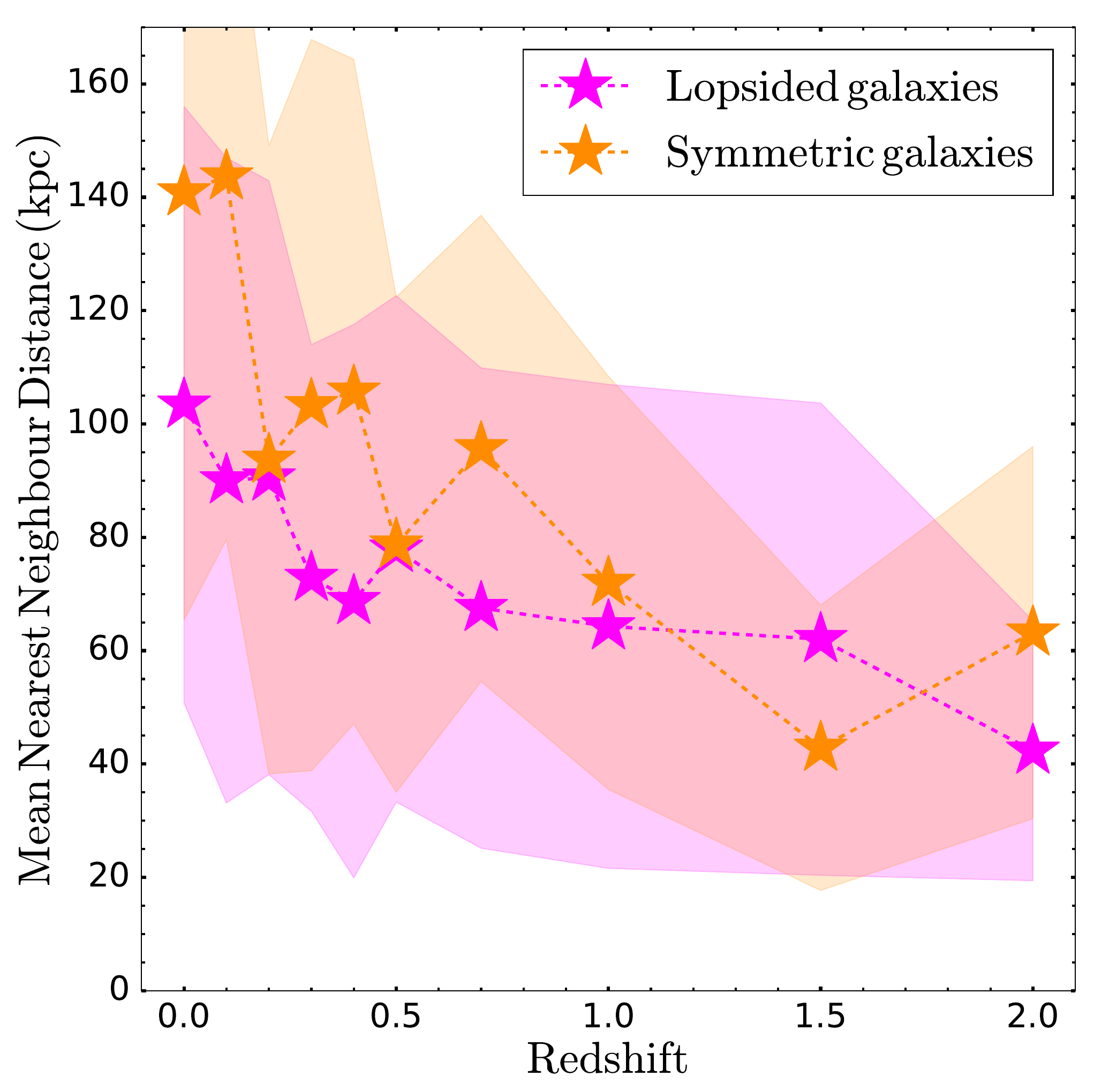}
    \caption{{\it Left panel:} Fraction of lopsided (magenta) and symmetric (orange) galaxies with one or more massive neighbors within $R_{200}$, whose stellar mass-ratio is $>1$:$10$. We normalize the number of lopsided and symmetric galaxies by the total number of lopsided and symmetric galaxies at each redshift between $0 < z < 2$, respectively. {\it Right panel:} Average distance to the closest massive neighbor (i.e. stellar mass-ratio is $>1$:$10$) of our galaxies, where the average is taken considering the sub-sample of the galaxies with one or more massive neighbors within $R_{200}$ at each redshift. Shaded areas are defined by the $25$th-$75th$ interquartile range. We consider here only the central galaxies in our samples.}
    \label{fig:merger_statistic}
\end{figure*}
 
\subsection{Origin of lopsidedness at high- and low-redshift}
\label{sec:origin_lopsidedness}
In the previous sections, we have seen that, while the internal properties of galaxies always play an important role for the development of lopsidedness, this perturbation seems to be significantly more common and stronger at high- than low-redshift. It is thus interesting to explore whether different mechanisms are at play, or whether the same mechanisms driving lopsidedness at $z\sim0$ are simply more efficient at high-redshift.

In this section, we study the relative importance as a function of redshift of different mechanisms that can potentially trigger lopsidedness, specifically focusing on close tidal interactions and gas accretion. We remind the reader that we are selecting a new galaxy sample at each redshift, using the same selection criteria described in Sec. \ref{sec:selected_disk_galaxies}. 

\subsubsection{Galaxy interactions}
\label{sec:mergers}
To study the role of tidal interactions as triggers of lopsidedness, we quantify the number of nearest neighbors, with stellar mass-ratio $>1$:$10$ with respect to the central galaxy, located within the virial radius of the galaxy $R_{200}$. 
We use this quantity as a proxy of the probability that the galaxy can experience a recent tidal interaction with neighboring galaxies. The adopted mass-ratio cut is chosen to include both major (mass-ratio $1$:$1$-$1$:$4$) and minor (mass-ratio $1$:$4$-$1$:$10$) mergers, which are expected to be massive enough to influence the morphology of the galaxy \citep{Bournaud2004,Gomez2017}.

In the left panel of Fig. \ref{fig:merger_statistic}, we show the number of lopsided and symmetric galaxies with one or more massive neighbors within $R_{200}$ normalized by the total number of lopsided and symmetric galaxies at each redshift between $0 < z < 2$, respectively.  
In the right panel of Fig. \ref{fig:merger_statistic}, we show the average distance to the closest massive neighbor of the galaxy, where the average is taken considering the sub-sample of the galaxies with one or more massive neighbors within $R_{200}$ at each redshift. As in the left panel, we divide our galaxies into lopsided and symmetric.
Here, we would like to remind the reader that we are looking at different galaxies as a function of redshift, since we are selecting our samples of disk-like galaxies independently at each redshift (see Sec. \ref{sec:sample_selection}).
Additionally, we only include here the results for the central galaxies and not for the satellite ones. This is because, by definition, satellites galaxies have at least one massive neighbor that they are interacting with, namely the central galaxy of the group or cluster to which they belong.

We see that the fraction of lopsided galaxies with one or more massive neighbors within $R_{200}$ decreases from $\sim20\%$ at high-redshift (i.e. $z\gtrsim1$) to $\sim10\%$ at $z=0$. At the same time, the average distance to the closest massive neighbor of the galaxies increases in a similar way for both lopsided and symmetric galaxies from high- to low-redshift. 
This indicates that close tidal interactions are more common at high- than low-redshift, and they can be a trigger of lopsided perturbations in at least $20\%$ of lopsided galaxies at high-redshift.
However, the fact that, towards low-redshift, the fraction of galaxies with massive nearby neighbors decreases indicates that the role of tidal interactions as mechanisms triggering lopsidedness becomes less relevant. Therefore, other mechanisms must be at play to generate lopsidedness at low-redshift.

While the results in Fig. \ref{fig:merger_statistic} show that the environment and external interactions can, indeed, be playing a more significant role in triggering lopsidedness at high- than low-redshift, they also indicate that tidal interactions alone cannot be the only mechanism at play in the origin of lopsidedness. In fact, at high-redshift, we see that $\sim80\%$ of lopsided galaxies do not have any close massive neighbors within $R_{200}$. This fraction increases to $\sim90\%$ at low redshift. This suggests that, in the majority of galaxies, lopsidedness must be mainly triggered through other mechanisms at all redshift between $0 < z < 2$. 
We note here that we are not considering the role of recent fly-bys in triggering lopsidedness. It can be that the galaxy experienced a recent fly-by event that triggered a lopsided perturbation, which is still observable after the companion has left the galaxy virial radius $R_{200}$. In fact, \citet{Bournaud2005} have shown that tidal interactions can produce strong lopsidedness lasting a few $\sim1$-$2\, \mathrm{Gyr}$.

For the symmetric galaxies, we see that the fraction of galaxies with one or more massive neighbors within $R_{200}$ is roughly constant at $10\%$ at all redshift. This means that, in addition to their internal properties that makes them less prone to develop lopsidedness (see Sec. \ref{sec:lopsidedness_galaxy_properties}), these galaxies have also not experienced significant interactions.

\begin{figure*}
    \centering
    \includegraphics[width=0.33\linewidth]{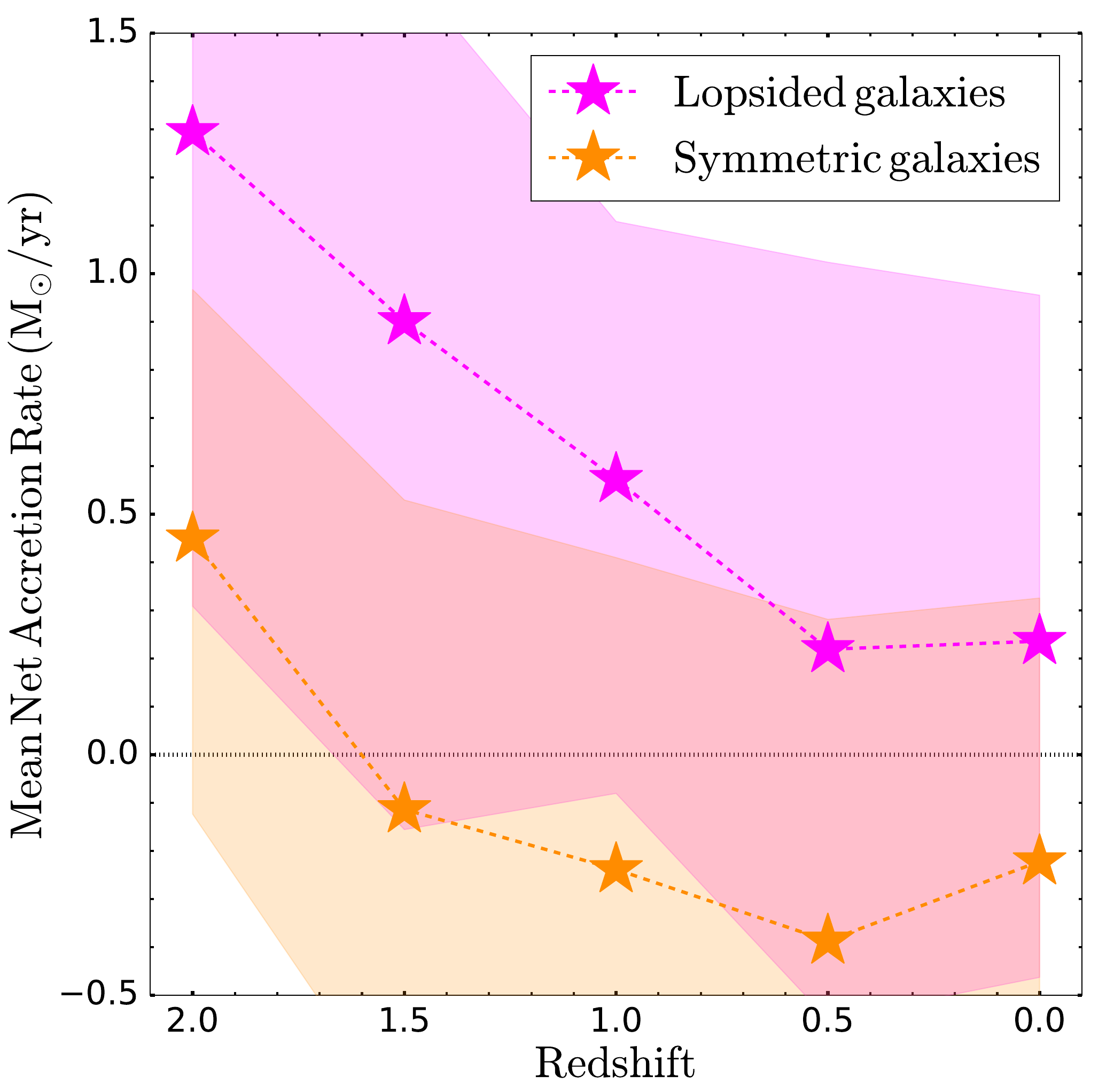}
    \includegraphics[width=0.33\linewidth]{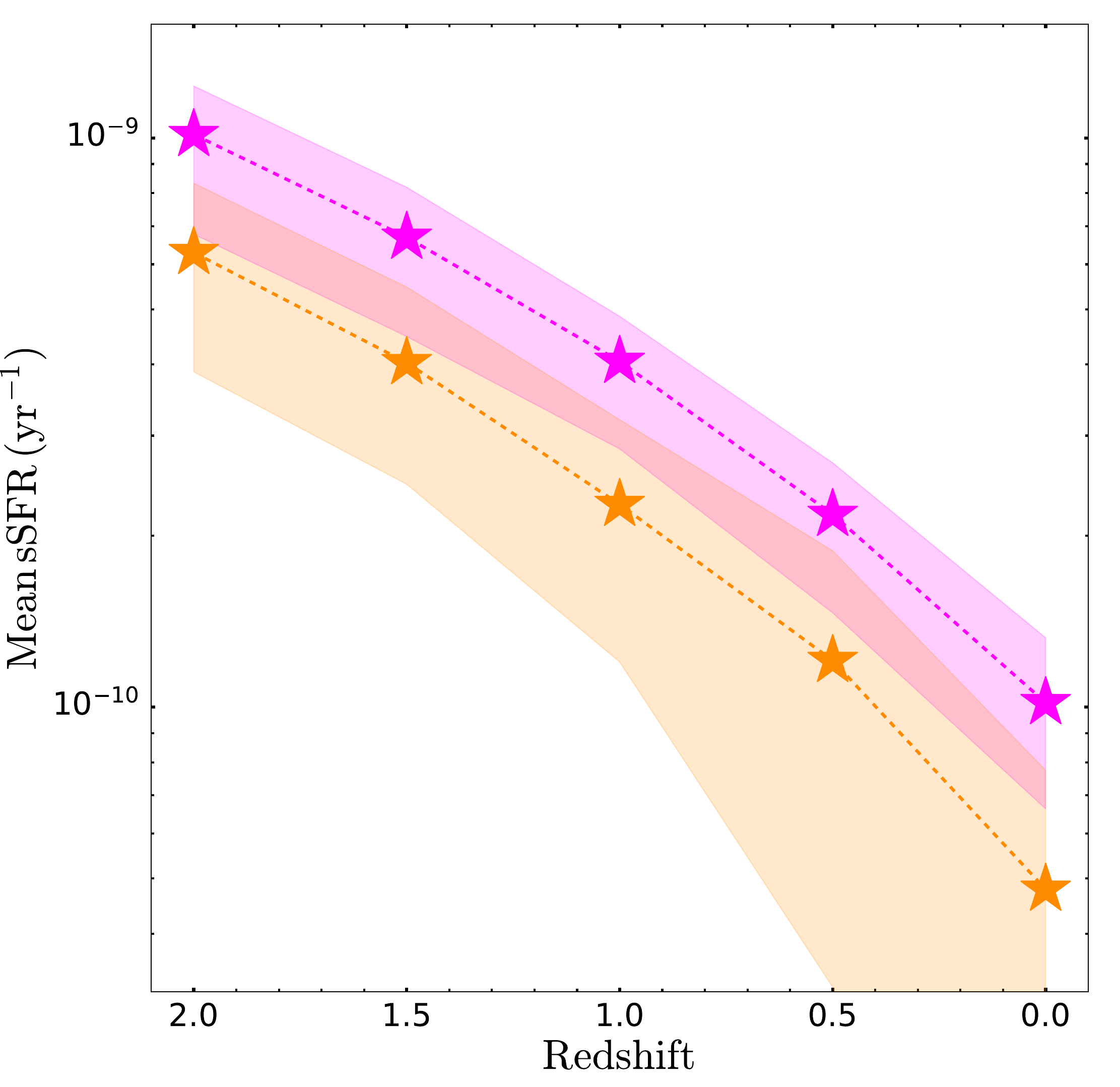}
    \includegraphics[width=0.33\linewidth]{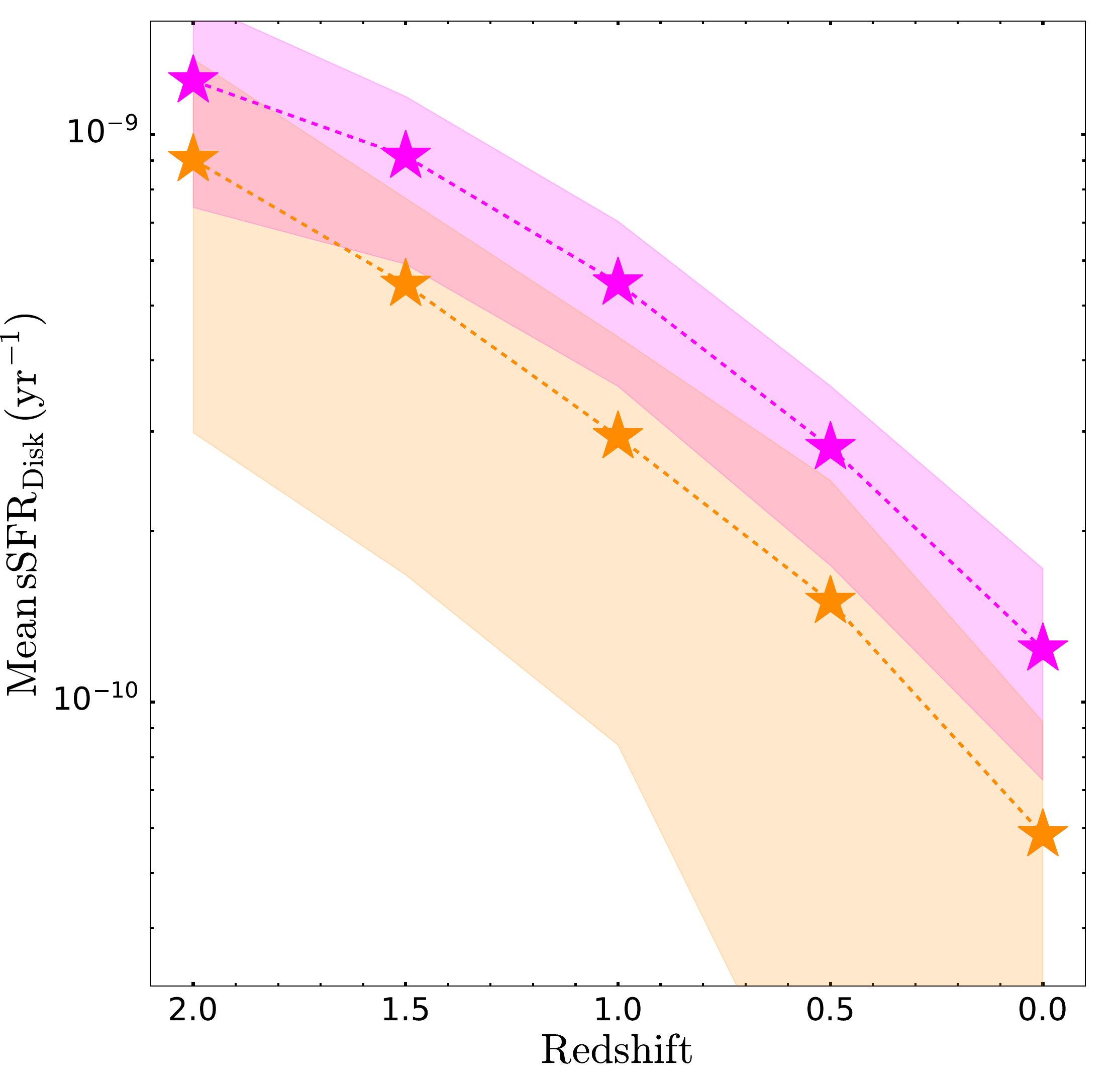}
    \caption{{\it Left panel:} Net accretion rate onto the galactic disk within the last $1\, \mathrm{Gyr}$ for our selected sample of disk-like galaxies at the five different specific redshift $z=2,\, 1.5,\, 1,\, 0.5\, \mathrm{and}\, 0$. {\it Middle panel:} Total galaxy star formation rate. {\it Right panel:} Star formation rate within the galactic disk, where the disk is defined as described in Sec. \ref{sec:bulge_and_disk}. We divide our galaxies into lopsided and symmetric, using the threshold $\mathrm{A}_{1} = 0.1$. Shaded areas are defined by the $25$th-$75$th interquartile range of the data at each redshift.}
    \label{fig:net_gas_accretion}
\end{figure*}

\subsubsection{Net accretion rate}
\label{sec:gas_accretion}
In this section, we explore the potential role of gas accretion as trigger of lopsidedness. For this purpose, we calculate net accretion rates of our samples of disk-like galaxies selected at the five different specific redshift $z=2,\, 1.5,\, 1,\, 0.5\, \mathrm{and}\, 0$. The net accretion rates are computed only considering a $1\, \mathrm{Gyr}$ time period by following the procedure described in \citet{Iza2022}: 

\begin{equation}
    \dot{\mathrm{M}}_{\mathrm{net}}(i) = \frac{ \mathrm{M}_{\mathrm{gas}}(i) - \mathrm{M}_{\mathrm{gas}}(i-1) + \mathrm{M}_{*} }{t(i) - t(i-1)},
\label{eq:net_accretion_rate}
\end{equation}

where $\mathrm{M}_{\mathrm{gas}}(i)$ is the total gas mass in the disk at snapshot $(i)$, while $\mathrm{M}_{*}$ is the total stellar mass born in the disk in the corresponding time interval $[t(i) - t(i-1)]$ considered for this analysis. We recall here that the disk is defined as the cylindrical region enclosed within the radius $2\, \mathrm{kpc} < r < R_{90}$ and vertical distance from the disk plane $-h_{90} < z < +h_{90}$, as described in Sec. \ref{sec:bulge_and_disk}.

In Fig. \ref{fig:net_gas_accretion}, from the left to the right panel, we show the net accretion rate, the galaxy specific star formation rate and the specific star formation rate within the disk as a function of redshift. As previously in Fig. \ref{fig:merger_statistic}, we divide our galaxies into lopsided and symmetric, using the threshold $\mathrm{A}_{1}=0.1$ as described in Sec. \ref{sec:lopsidedness_evolution}. 

In the left panel of Fig. \ref{fig:net_gas_accretion}, we see that lopsided galaxies are characterized by larger net accretion rate than symmetric galaxies at all redshift. In particular, for lopsided galaxies, we see that the net accretion rate decreases from $z=2$ to $z=0$, indicating that these galaxies experience more significant gas accretion at high- (i.e. $z>1$) than low-redshift. 
On the contrary, symmetric galaxies are mainly characterized by negative net accretion rate, meaning either that they experienced more significant outflow events that removed the gas from the disk, or that they consumed more gas than have accreted during the last $\sim10\, \mathrm{Gyr}$ of evolution.

In the middle and right panel of Fig. \ref{fig:net_gas_accretion}, we also see that lopsided galaxies are characterized by, on average, larger specific star formation rate than symmetric ones at all redshift both globally and within the disk region.   
This suggests that the gas accreted onto the disk is being more efficiently converted into stars through subsequent star formation in lopsided galaxies. 
We note that we find similar results when we consider central and satellite galaxies, separately.

Overall, the results in Fig. \ref{fig:net_gas_accretion} show that gas accretion with subsequent star formation is clearly more prominent for lopsidedness at all redshift between $0 < z < 2$. The fact that the net accretion rate decreases from high- to low-redshift also suggests that the role of gas accretion behind lopsidedness is more significant at high- than low-redshift, consistent with the decreasing fraction of lopsided galaxies towards low-redshift, as shown in Fig. \ref{fig:lopsidedness_evolution}.
In a previous work, \citet{Lokas2022} also suggested that asymmetric gas accretion followed by star formation is a probable mechanism for the origin of lopsidedness at $z=0$, based on the finding that lopsided galaxies are typically characterized by larger gas fraction, larger star formation rate, lower metallicity and bluer color than symmetric galaxies, using the IllustrisTNG simulations. In another work, \citet{Bournaud2005} simulated an ideal case where the accretion of gas onto the galaxy occurs along one off-centered filament or few filaments. They found that this asymmetric gas accretion can produce strong lopsided amplitudes. However, they did not try to study more realistic gas accretion scenarios predicted by cosmological models.
We note that, here, we have not studied whether the gas is being accreted asymmetrically nor its angular momentum. The latter is also expected to be an important parameter due to the recent finding that lopsided galaxies tend to live in high-spin halos \citep{Varela-Lavin2023}, which can lead to the formation of their more extended disks \citep{Grand2017}. We will perform this analysis in details in a follow-up work to test the scenario of lopsidedness being generated as a result of asymmetric gas accretion with subsequent star formation in realistic cosmological models.

\begin{figure}
    \centering
    \includegraphics[width=0.45\textwidth]{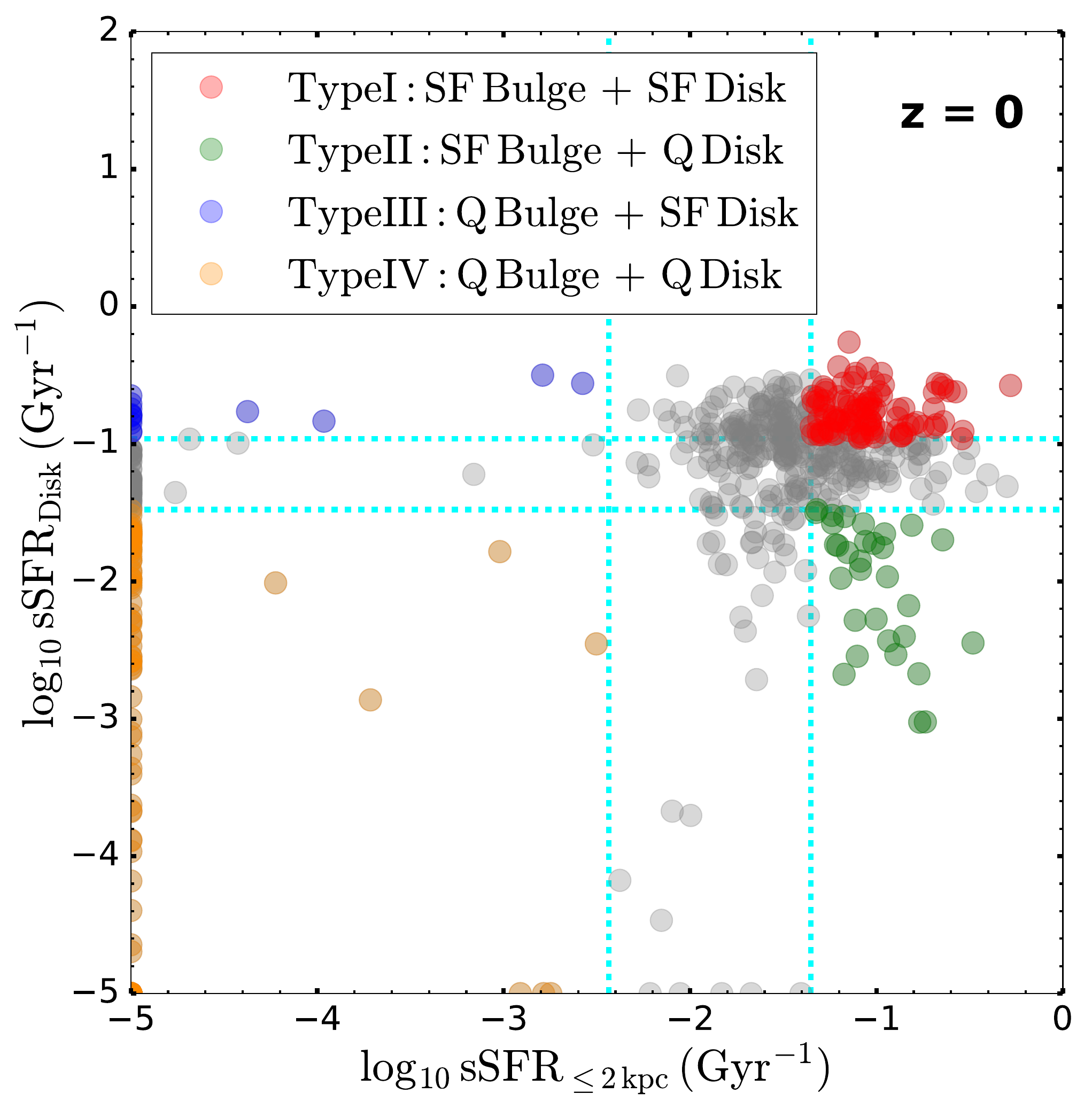}
    \includegraphics[width=0.45\textwidth]{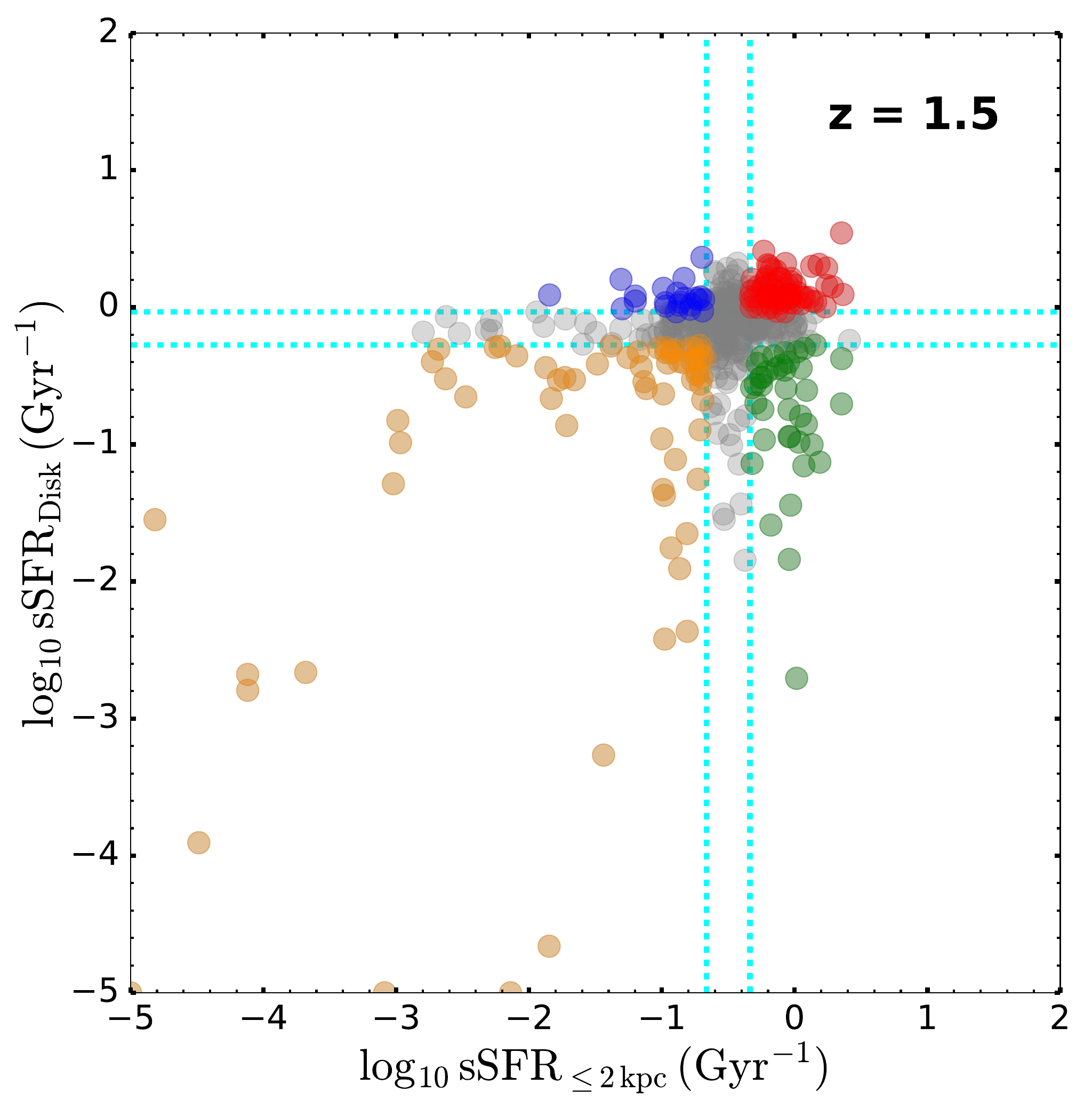}
    \caption{Distribution of the specific star-formation rate (sSFR) of the central regions and disks of our selected sample of disk-like galaxies at $z=0$ (top panel) and $z=1.5$ (bottom panel). The sSFR of the different components of the galaxies are calculated as described in Sec. \ref{sec:star_formation_rates_classification}. We divide the sSFR of the central regions and disks in tertiles (dashed lines) and we select the four galaxy sub-samples lying on the outside of the tertile contours (coloured points). These four galaxy sub-samples represent the sub-samples of star-forming cores (disks) and quenched cores (disks) defined in Sec. \ref{sec:star_formation_rates_classification}, similarly to \citet{LeBail2023}.}
    \label{fig:sSFR_galaxy_classification}
\end{figure}

\section{Comparison with observations}
\label{sec:comparison_observations}
In a recent work, \citet{LeBail2023} investigated the properties of a small sample of $22$ dusty star-forming galaxies observed with the JWST Near Infra-Red Camera between $1.5 < z < 3$. They found that $64\%$ of their galaxies are lopsided. Specifically, they find that the sub-sample of strongly lopsided galaxies tend to be characterized by a star-forming core with low core mass fraction compared to the symmetric sub-sample, which is typically characterized by a quenched bulge with high core mass fraction (see figure 14 in \citet{LeBail2023}).
Based on these results, \citet{LeBail2023} suggest that the build-up of a quenched massive bulge component can have the effect of stabilizing the disk against the lopsided perturbation, consistent with the results from previous works at $z=0$ that demonstrated that the low central stellar mass densities are key for the onset of lopsided perturbations \citep{Reichard2008,Varela-Lavin2023,Dolfi2023}.
In this section, we propose to perform a comparison with the observations from \citet{LeBail2023} by classifying our selected sample of simulated disk-like galaxies into different types based on their star formation activity, similarly to \citet{LeBail2023}.
We remind the reader that we are selecting a new galaxy sample at each redshift, using the same selection criteria described in Sec. \ref{sec:selected_disk_galaxies}. 

\begin{figure}
    \centering
    \includegraphics[width=0.45\textwidth]{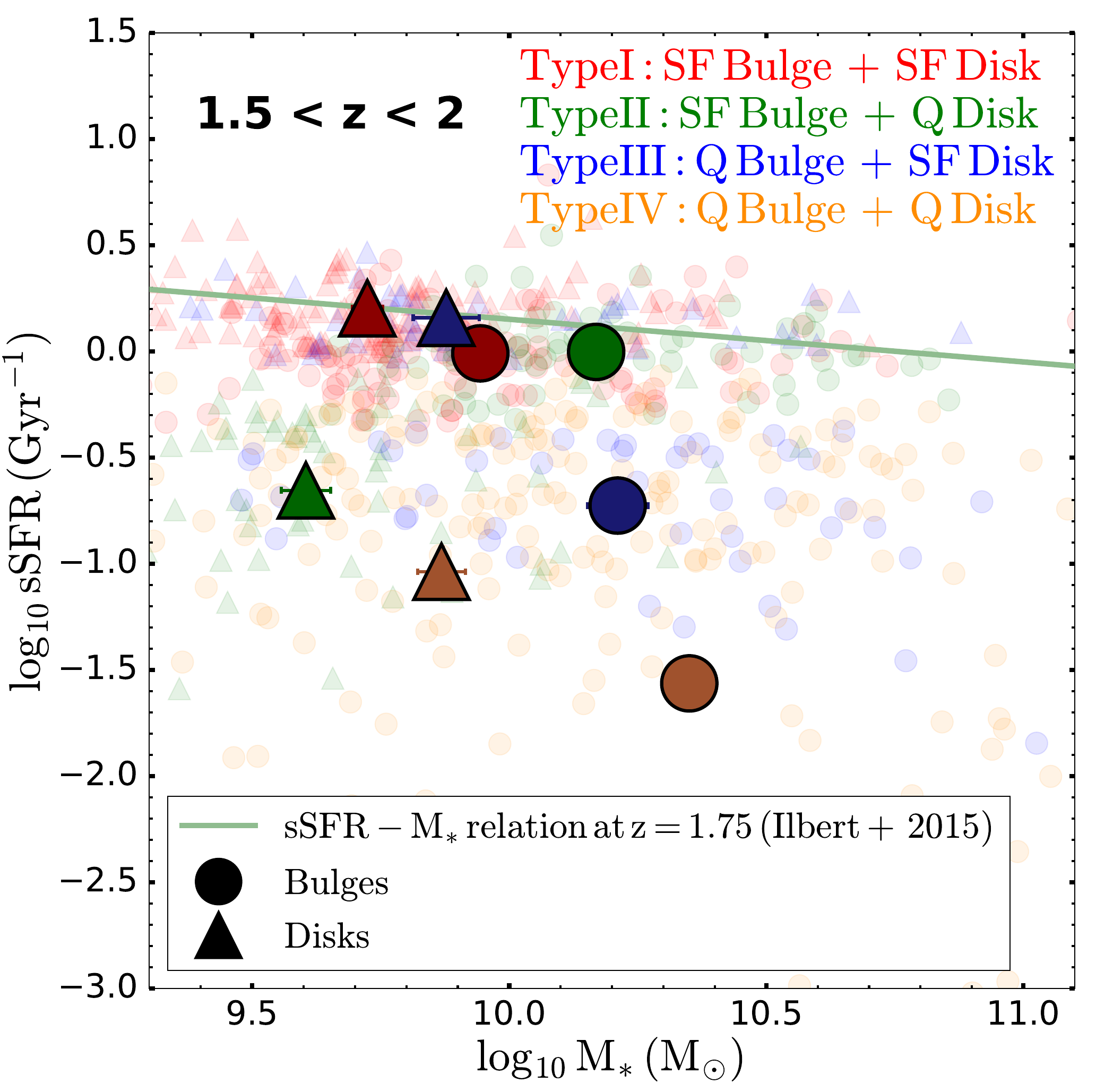}
    \includegraphics[width=0.45\textwidth]{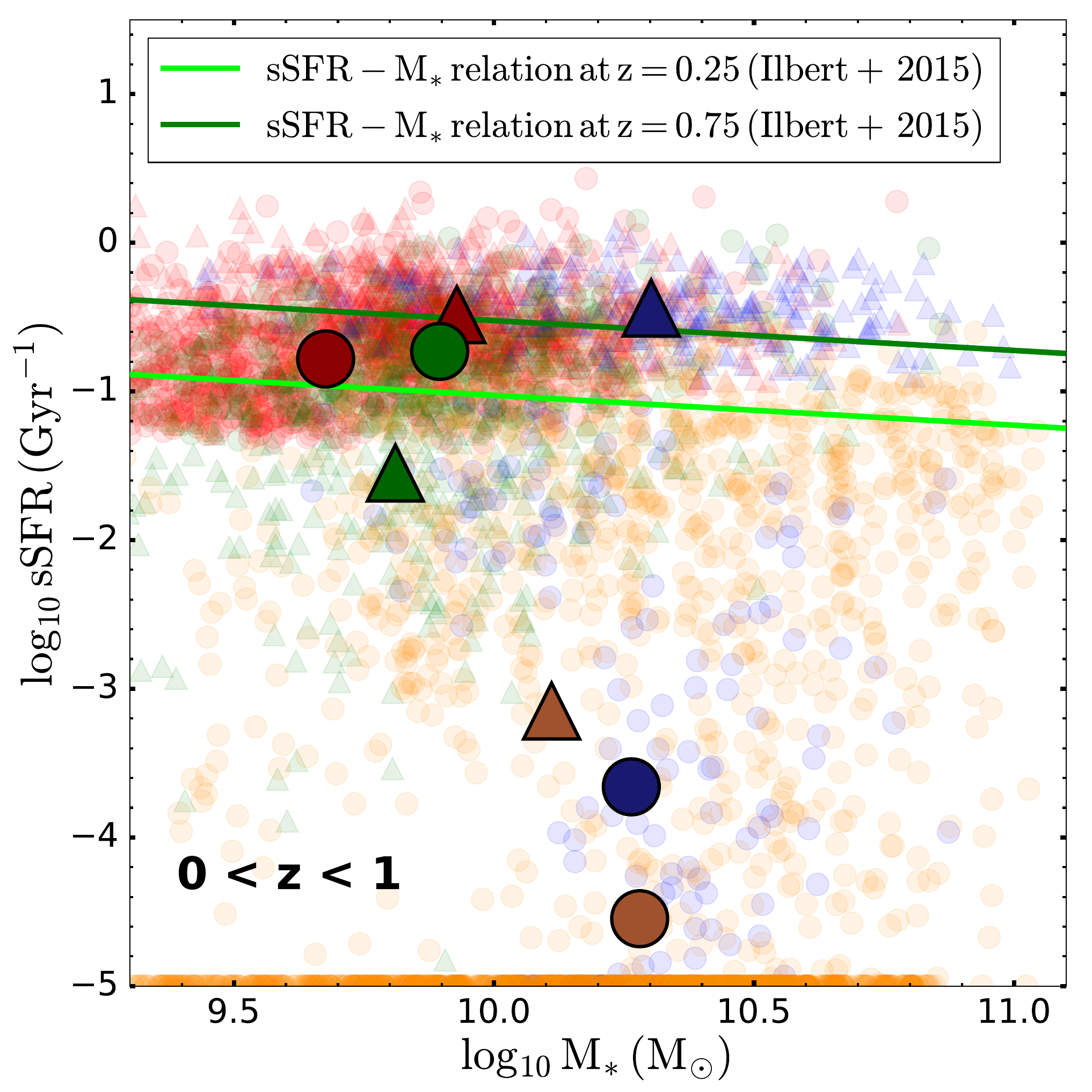}
    \caption{Specific star-formation rate (sSFR)-stellar mass relation of the central regions and disk components of our selected sample of disk-like galaxies between $0 < z < 2$, divided into the four galaxy types defined in Fig. \ref{fig:sSFR_galaxy_classification} and in the two redshift intervals: $1.5 < z < 2$ (i.e. high-redshift) and $0 < z < 1$ (i.e. low-redshift). Small symbols represent the sSFR and stellar mass of the central regions and disk components of individual galaxies, while large symbols represent the mean values of the sSFR and stellar mass of the central regions and disk components for the galaxies belonging to each type. Errorbars represent the standard error of the mean. The light-to-dark green solid lines represent the redshift-dependent parametrization of the star-forming main-sequence (SFMS) relation defined in equation 5 of \citet{Ilbert2015}.}
    \label{fig:ssfr_mass_relation}
\end{figure}

\subsection{Classification: star-formation rate of cores and disks}
\label{sec:star_formation_rates_classification}
Firstly, we calculate the star-formation rate (SFR) of the central and disk components of each one of our selected disk-like galaxies between $0 < z < 2$, as described in Sec. \ref{sec:bulge_and_disk}. Then, we calculate the specific star-formation rate (sSFR) of the central and disk components of our galaxies by normalizing the SFR of each component by its total stellar mass.

In Fig. \ref{fig:sSFR_galaxy_classification}, we show the distribution of the sSFR of the central and disk components of our selected sample of disk-like galaxies at $z=0$ and $z=1.5$, as an example. 
At each redshift, we separate the distribution of the sSFR of both components in tertiles (dashed lines in Fig. \ref{fig:sSFR_galaxy_classification}) and we select the four galaxy sub-samples lying on the outside of the tertile contours (colored points in Fig. \ref{fig:sSFR_galaxy_classification}). We define the following four galaxy types, similarly to \citet{LeBail2023}:

\begin{itemize}
    \item TypeI: sub-sample of galaxies characterized by both a star-forming core and a star-forming disk (SF bulge$+$SF disk, hereafter);
    \item TypeII: sub-sample of galaxies characterized by a star-forming core and a quenched disk (SF bulge$+$Q disk, hereafter);
    \item TypeIII: sub-sample of galaxies characterized by a quenched core and a star-forming disk (Q bulge$+$SF disk, hereafter);
    \item TypeIV: sub-sample of galaxies characterized by both a quenched core and a quenched disk (Q bulge$+$Q disk, hereafter). 
\end{itemize}

\begin{figure*}
    \centering
    \includegraphics[width=0.33\textwidth]{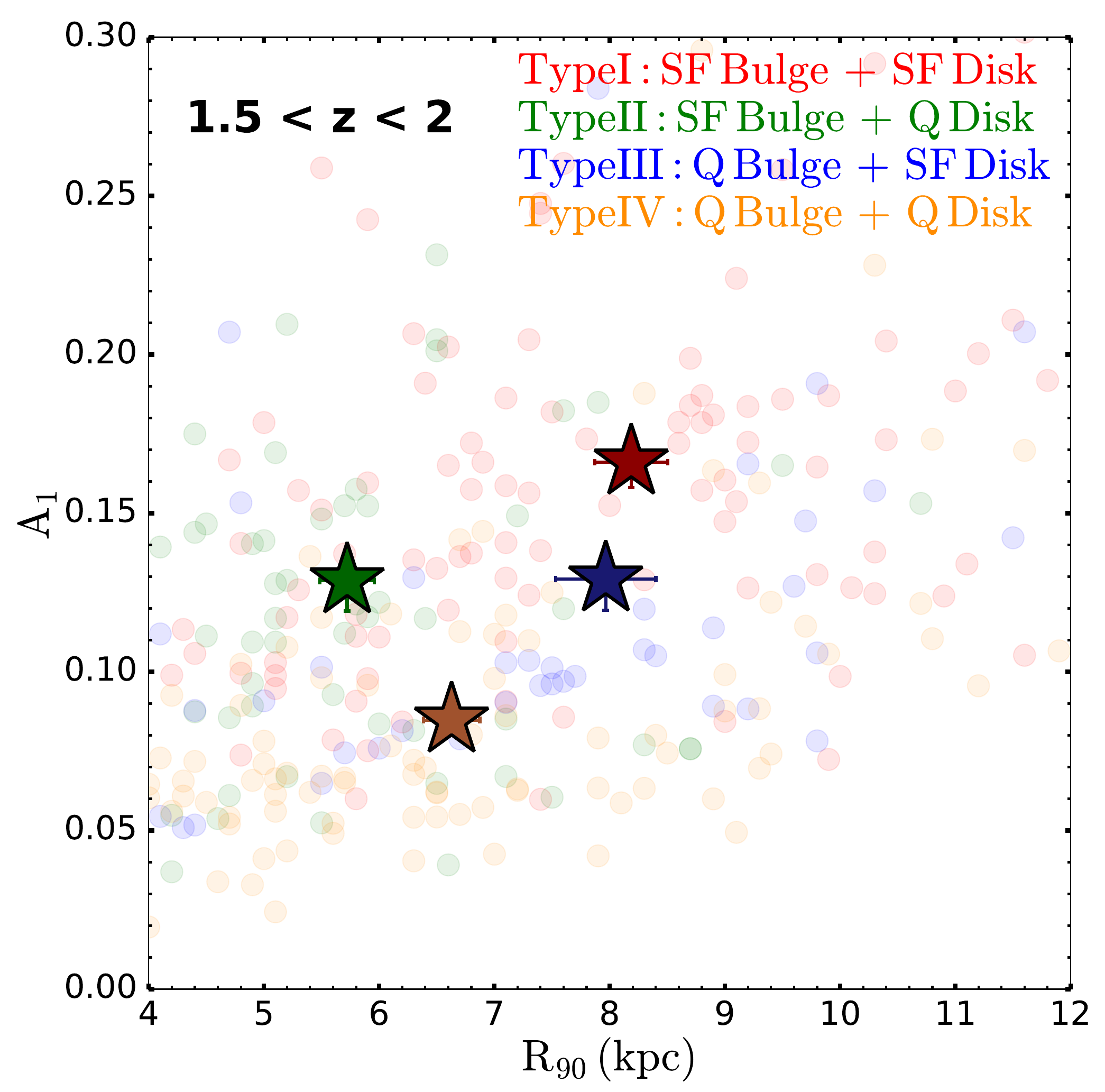}
    \includegraphics[width=0.33\textwidth]{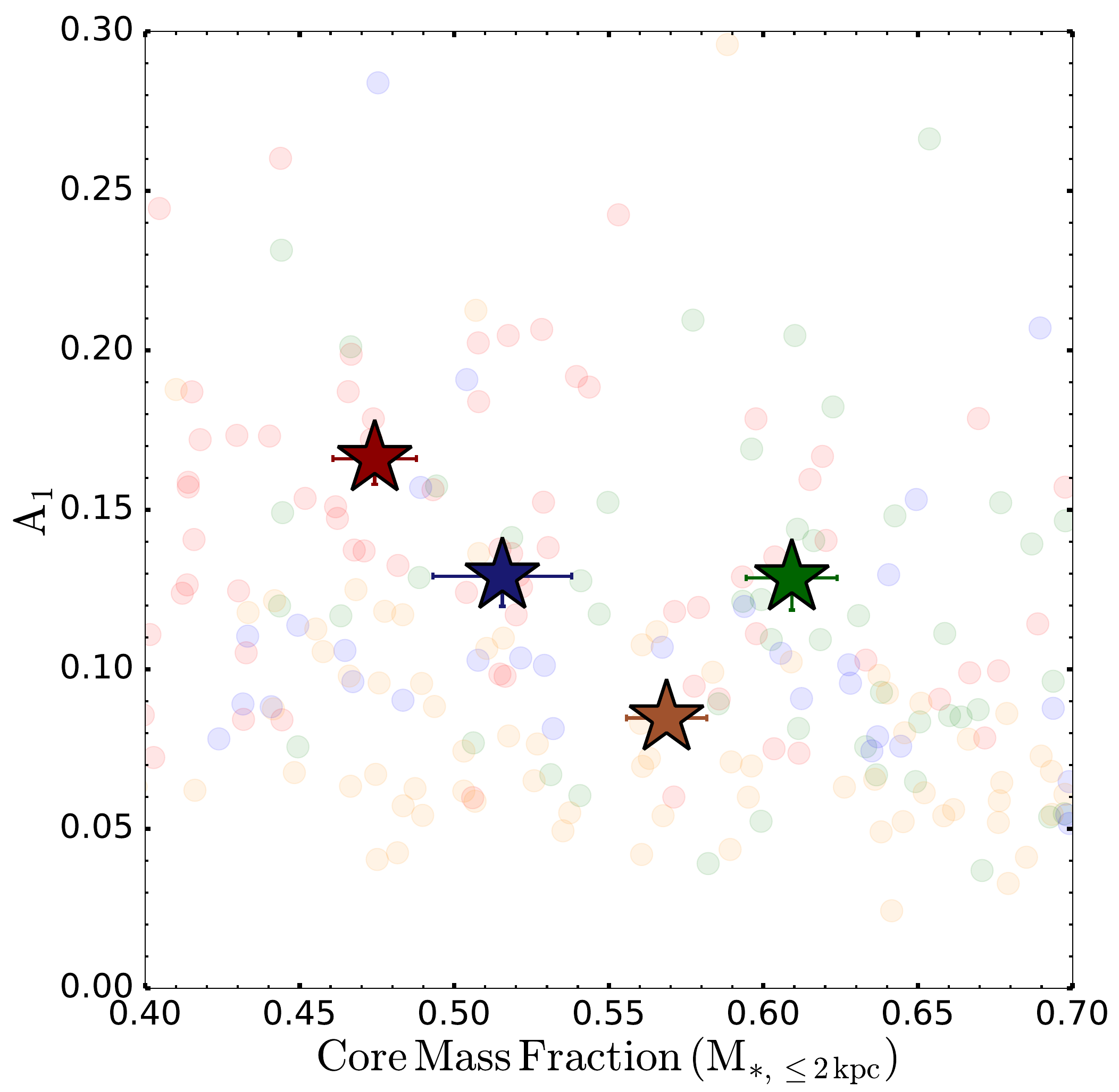}
    \includegraphics[width=0.33\textwidth]{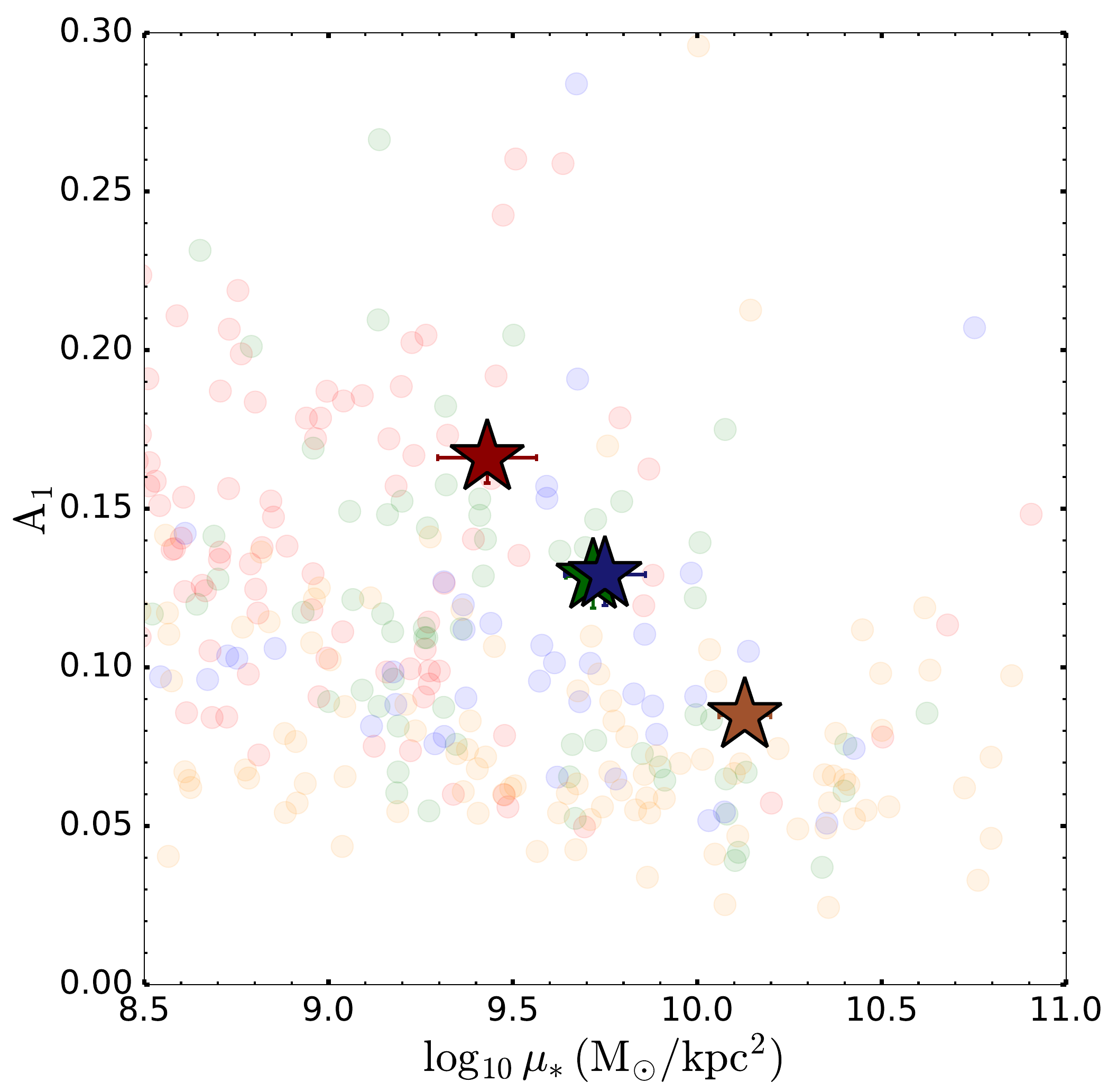}

    \includegraphics[width=0.33\textwidth]{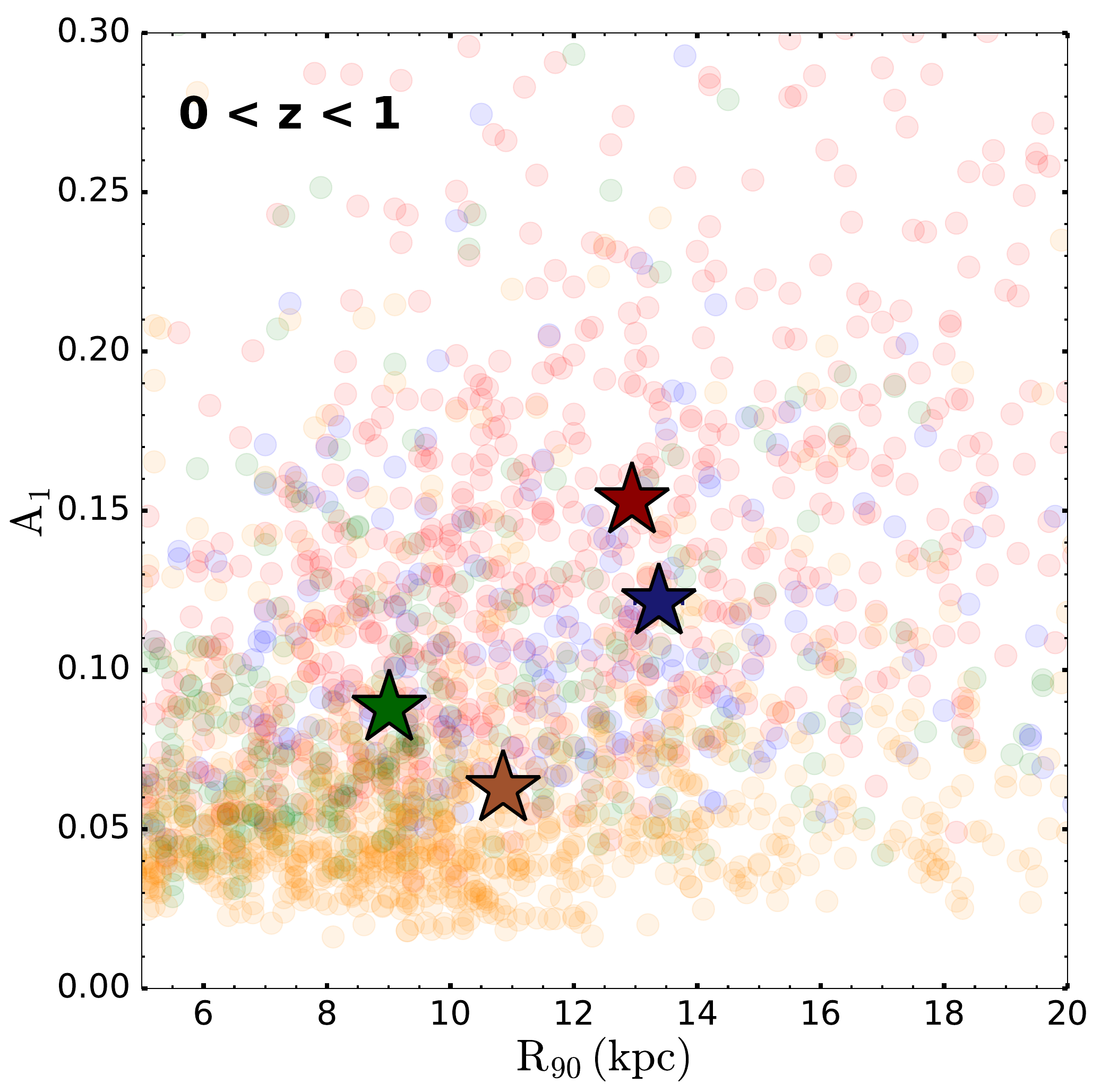}
    \includegraphics[width=0.33\textwidth]{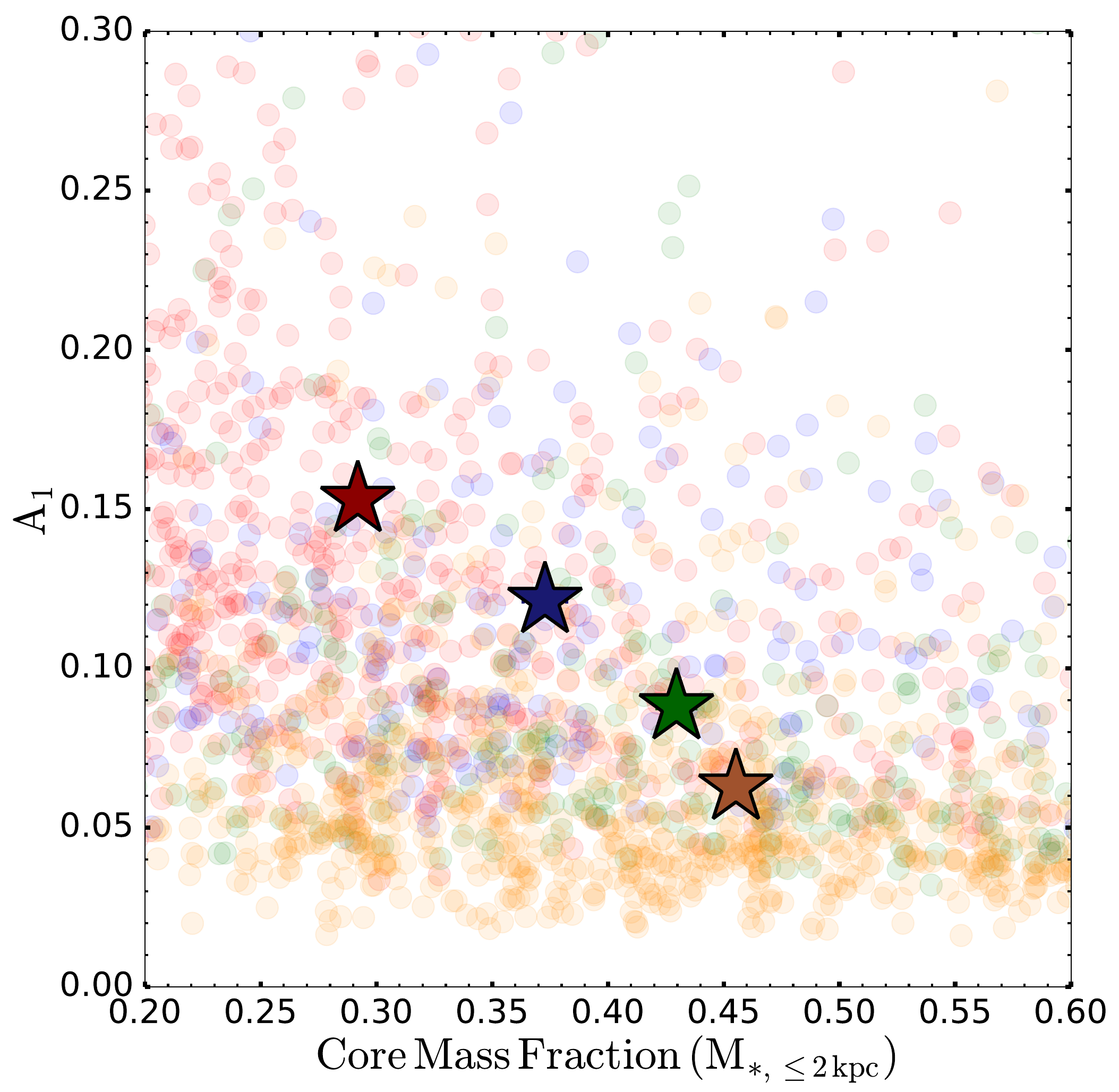}
    \includegraphics[width=0.33\textwidth]{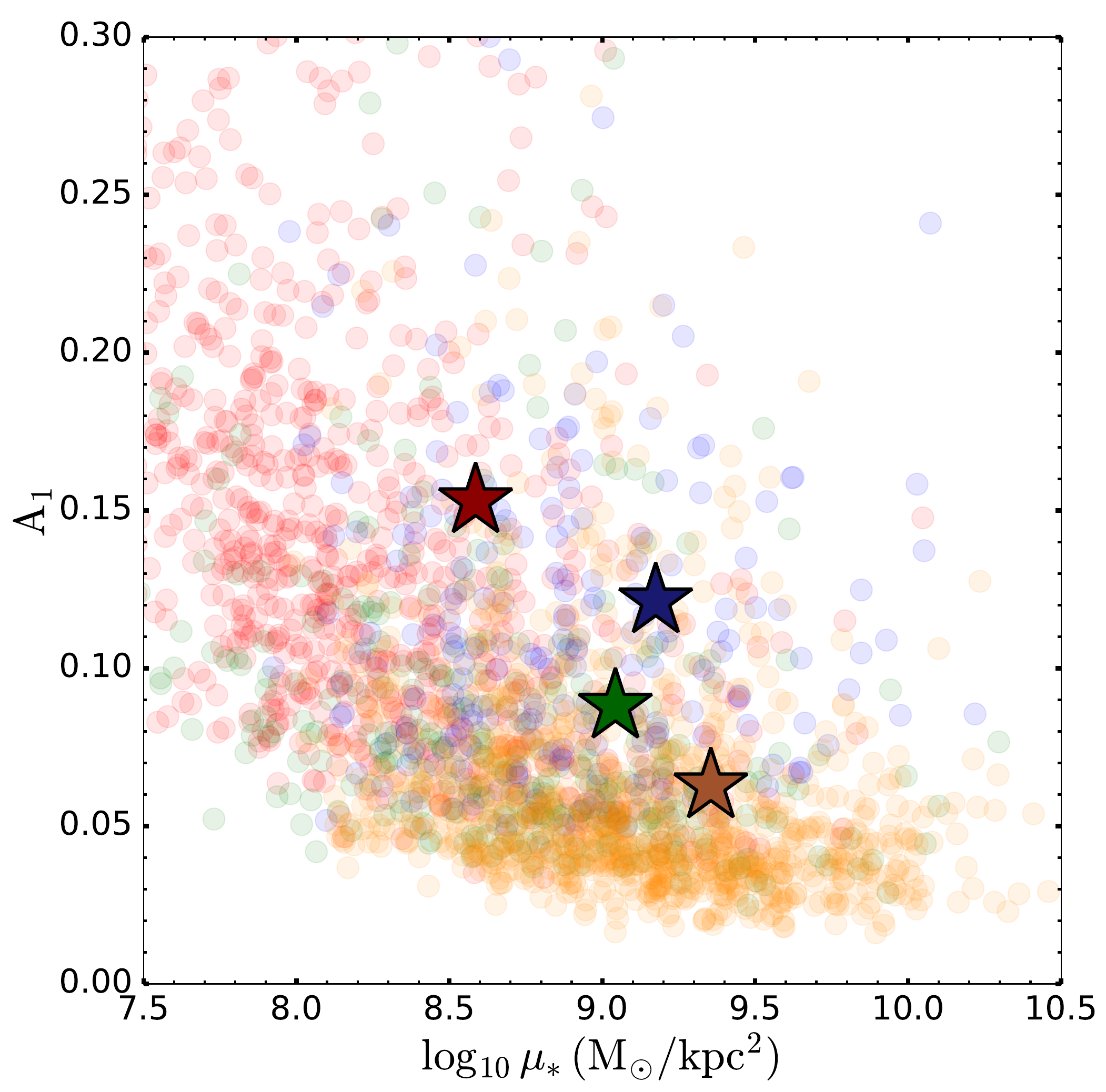}
    \caption{From the left to the right panel, the lopsided amplitude as a function of the disk size ($R_{90}$), fraction of stellar mass within the central regions ($\mathrm{M}_{\mathrm{*}}(\leq2\, \mathrm{kpc})/\mathrm{M_{*}}$) and central stellar mass density ($\mu_{*}$), respectively, for our selected sample of disk-like galaxies between $0 < z < 2$, divided into the four galaxy types defined in Fig. \ref{fig:sSFR_galaxy_classification} and in the two redshift intervals: $1.5 < z < 2$ (i.e. high-redshift) and $0 < z < 1$ (i.e. low-redshift). Small symbols represent individual galaxies, while large symbols represent the mean distribution of each galaxy type. Errorbars represent the bootstrap standard error, calculated from the standard deviation of the means of $500$ bootstrap samples obtained by sampling with replacement our original dataset.}
    \label{fig:asymmetry_vs_galaxy_properties}
\end{figure*}

We note that the term "quenched" here does not necessarily mean that the central regions or disks are completely quiescent, but they can still be mildly star-forming according to our classification. We apply the same classification shown in Fig. \ref{fig:sSFR_galaxy_classification} to all our sample of disk-like galaxies selected at each redshift between $0 < z < 2$.
We note that our galaxy classification into TypeI-TypeIV differs from that of \citet{LeBail2023}. In this work, we use the instantaneous star formation rate of the gas cells to calculate the total SFR of each galaxy component. \citet{LeBail2023} estimated the SFR, stellar mass and magnitudes of each galaxy component by fitting its observed SED. 
Then, they separated each galaxy component into quiescent and star-forming based on its location on the $UVJ$ color diagram. Creating mock observations to reproduce the galaxy classification from \citet{LeBail2023} is outside the scope of this paper. Nonetheless, the selection criteria applied in this work allows a first and reasonable qualitative comparison with the observational results of \citet{LeBail2023}.

\subsection{Star-forming main sequence}
\label{sec:ssms}
In Fig. \ref{fig:ssfr_mass_relation}, we show the sSFR-stellar mass relation of the central and disk components of our selected sample of disk-like galaxies between $0 < z < 2$. We divide our galaxies into the four galaxy types defined in Sec. \ref{sec:star_formation_rates_classification}. Small symbols represent the sSFR and stellar mass of the central and disk components of individual galaxies, while large symbols represent the average sSFR and stellar mass of the central and disk components for the galaxies belonging to each type. Solid lines represent the redshift-dependent parametrization of the star-forming main-sequence (SFMS) relation defined in equation 5 of \citet{Ilbert2015}. Here, and throughout the rest of the paper, we will separate our galaxies in two redshift intervals: $1.5 < z < 2$ (i.e. high-redshift) and $0 < z < 1$ (i.e. low-redshift), where the high-redshift interval partly overlaps with the redshift range of the observations from \citet{LeBail2023}.
This figure shows that the mean values of the sSFR of the star-forming central and disk components of TypeI, TypeII and TypeIII galaxies are consistent with the SFMS of the corresponding high- and low-redshift intervals. On the other hand, we see that the quenched core and disk components of TypeII, TypeIII and TypeIV galaxies all lie below the SFMS of the corresponding high- and low-redshift intervals. These results are in agreement with our classification of the galaxies into star-forming (quenched) cores and/or disks, described in Sec. \ref{sec:star_formation_rates_classification}. We note however that, in the low-redshift interval, the quenched disks of TypeII galaxies have, on average, higher sSFR than the other quenched components. 

\subsection{Lopsidedness}
\label{sec:asymmetry}
In Fig. \ref{fig:asymmetry_vs_galaxy_properties}, we show the correlation between lopsidedness and the internal properties of our selected sample of disk-like galaxies between $0 < z < 2$. 
We focus here on the correlation between lopsidedness and disk size ($R_{\mathrm{90}}$), fraction of stellar mass within the inner $2\, \mathrm{kpc}$ ($\mathrm{M}_{\mathrm{*,\leq2\, kpc}}/\mathrm{M_{*}}$; core mass fraction, hereafter) and central stellar mass density ($\mu_{*}$) within the inner $R_{\mathrm{h}}$. 

For the high-redshift interval, we find a clear correlation between the lopsided amplitude and central stellar mass density. Indeed, the top-right panel shows $\mathrm{A}_{1}$ decreasing with increasing $\mu_{*}$ from TypeI to TypeIV galaxies. TypeII and TypeIII galaxies, which are characterized by similar central stellar mass density, show similar lopsided amplitude.
The core mass fraction (top-middle panel) and disk size (top-left panel) also decrease from TypeI to TypeIV galaxies. However, the TypeII and TypeIII galaxies, which have similar $\mathrm{A}_{1}$, differ in these quantities such that TypeII galaxies have smaller disk size and larger core mass fraction compared to the other galaxy types. 
Overall, these results confirm the finding that the central stellar mass density is one of the most important parameters for determining the strength of the lopsided amplitude at high-redshift, with the disk size and core mass fraction playing a secondary role.

For the low-redshift interval, we see that TypeI and TypeIII galaxies have, on average, larger lopsided amplitude than TypeII and TypeIV galaxies. However, while we see that the central stellar mass density and the strength of the lopsided amplitude show a very strong correlation, we see that the disk size and core mass fraction are also strongly correlated to $\mathrm{A}_{1}$. In fact, we see that TypeIII galaxies have, on average, larger lopsided amplitude than TypeII galaxies for similar (or slightly larger) central stellar mass density, but TypeIII galaxies tend to be characterized by lower core mass fraction and larger disk size than TypeII galaxies. In particular, TypeIII galaxies have similarly extended star-forming disks to TypeI galaxies. 
This suggests that lopsidedness in TypeI and TypeIII galaxies could be connected to the properties of their star-forming disk component, indicating that this perturbation may be associated with gas accretion at low-redshift, as previously discussed in Sec. \ref{sec:origin_lopsidedness}. 

We note that our results in Fig. \ref{fig:asymmetry_vs_galaxy_properties} show some differences with respect to the observational results from \citet{LeBail2023}. 
Specifically, \citet{LeBail2023} found that TypeIII galaxies have, on average, the highest core mass fraction and lowest lopsided amplitude (see their figure 14). 
Additionally, \citet{LeBail2023} also found that TypeIII galaxies have lower overall sSFR and are located at lower redshift than TypeI and TypeII galaxies (see their figure 12). 
For this reason,  \hbox{\citet{LeBail2023}} suggest that the TypeIII galaxy population is more evolved compared to the TypeI and TypeII galaxy populations and, thus, had more time to build up a massive quenched bulge component that can make the galaxy less susceptible to develop strong lopsidedness.  
On the contrary, in this work, we find that TypeIII galaxies at high-redshift have low core mass fraction, more similarly to TypeI galaxies, and can show large lopsided amplitude (see top-middle panel in Fig. \ref{fig:asymmetry_vs_galaxy_properties}).
This discrepancy between the observational results and our models may be due to several reasons. Firstly, it could be a result of the different methods adopted in this work to define the central and disk components of galaxies, as well as to measure the sSFR of each galactic component, with respect to the observations of \citet{LeBail2023}. This can lead to differences in the estimation of the internal properties of galaxies (e.g. stellar mass, central mass density) between observations and simulations. Secondly, it could be due to the different methods used to divide our galaxies into the four galaxy types with respect to \citet{LeBail2023} (see Sec. \ref{sec:star_formation_rates_classification}), as well as to the different methods used to classify galaxies into lopsided and symmetric in the observations of \citet{LeBail2023} and our simulations. In fact, \citet{LeBail2023} quantified lopsidedness by rotating the galaxy image by $180\degr$ and subtracting it from the original image, while we quantify lopsidedness via Fourier analysis as described in Sec. \ref{sec:lopsidedness}. Finally, it could also be a result of the different redshift interval covered by our simulated galaxy models (i.e. $1.5 < z < 2$) and observed galaxies from \citet{LeBail2023} (i.e. $1.5 < z < 3$) at high-redshift. 

\begin{figure*}
    \centering
    \includegraphics[width=0.45\textwidth]{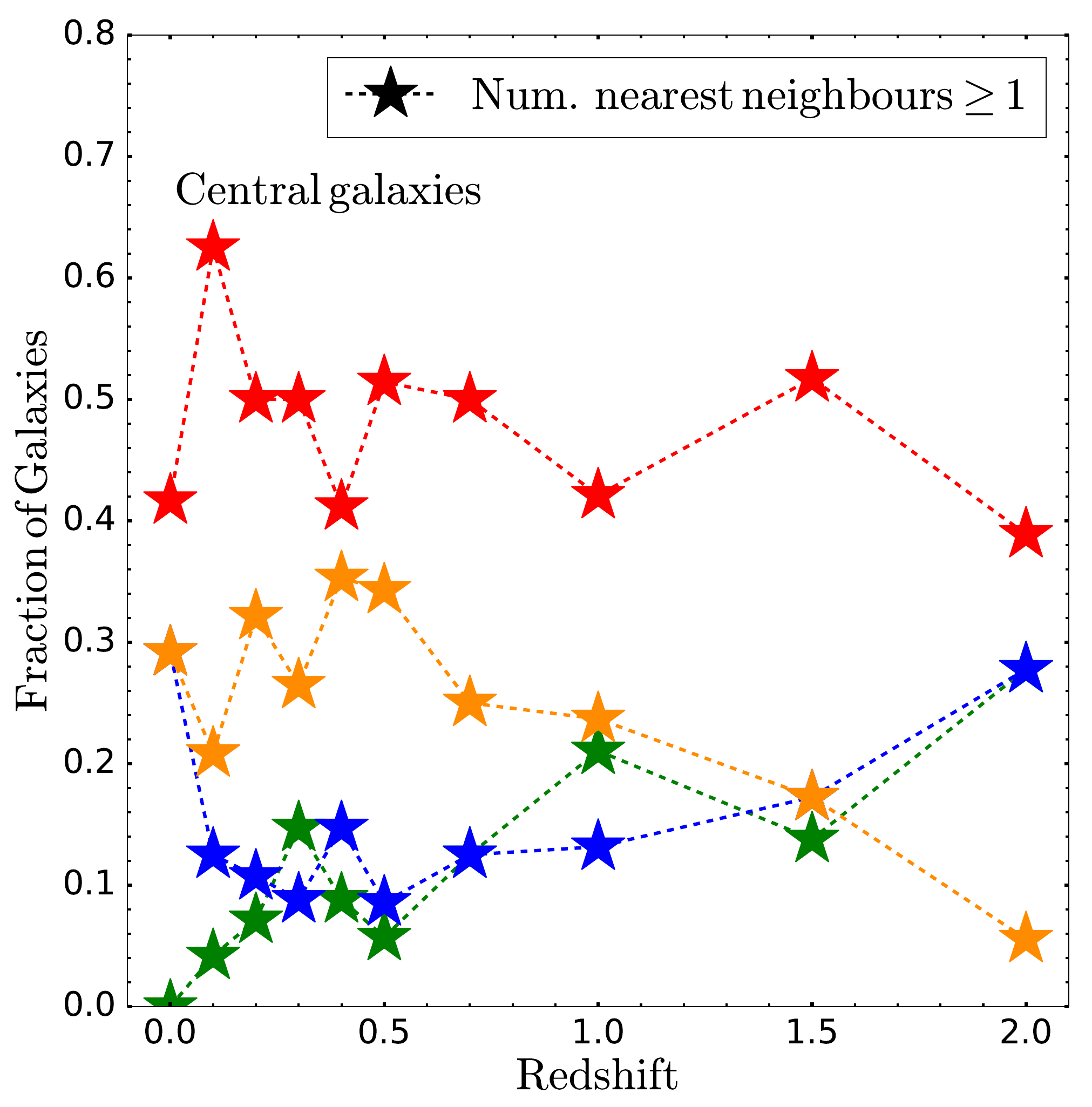}
    \includegraphics[width=0.45\textwidth]{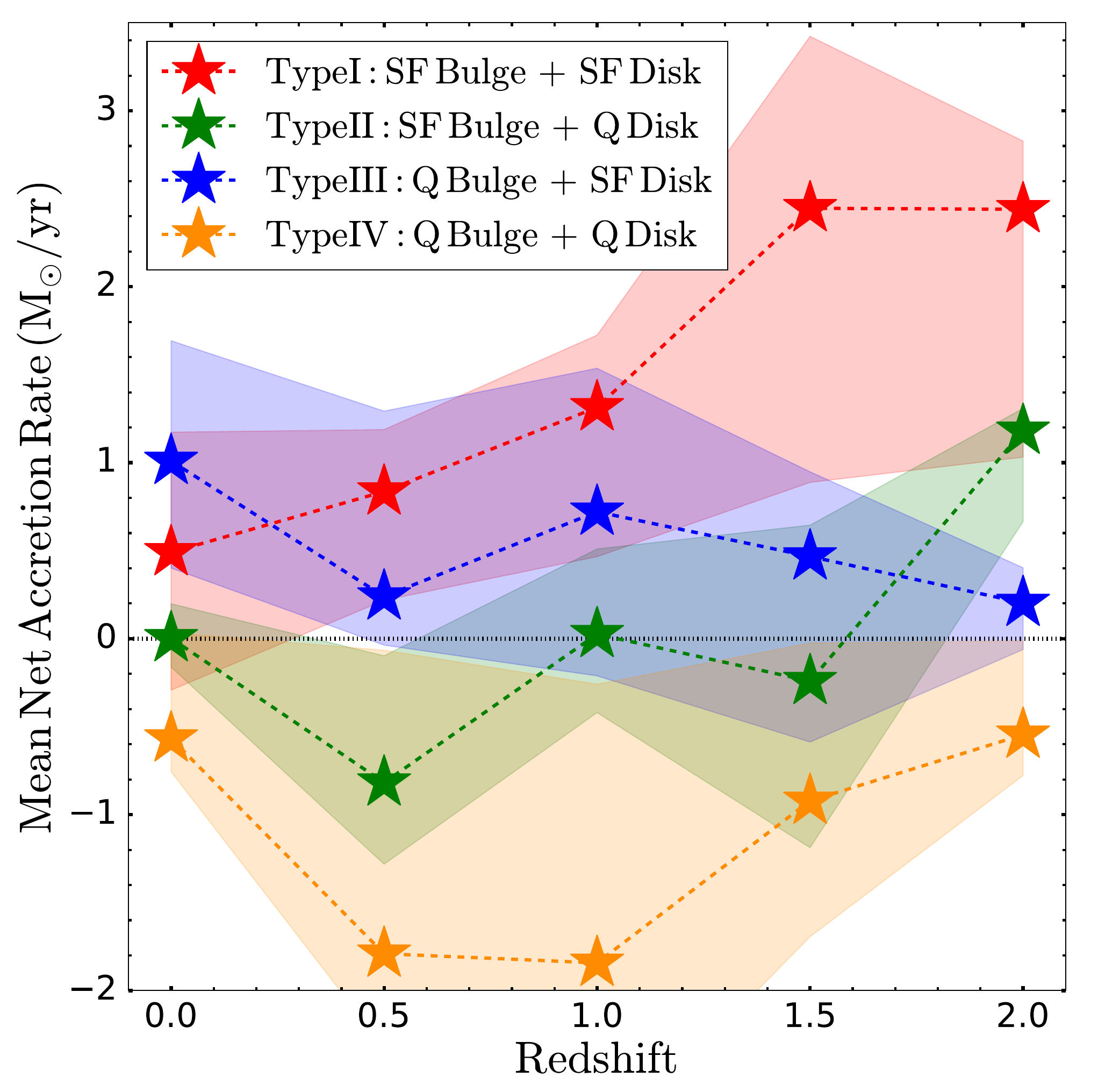}
    \caption{{\it Left panel:} Fraction of galaxies with one or more massive neighbors with stellar mass-ratio $>1$:$10$ within $R_{200}$ as a function of redshift between $0 < z < 2$. Here, we show only the sub-sample of the central galaxies with at least one nearest neighbor. {\it Right panel:} Mean net accretion rate onto the galactic disk within the last $1\, \mathrm{Gyr}$ of our sample of disk-like galaxies selected at the five different specific redshift $z=2,\, 1.5,\, 1,\, 0.5\, \mathrm{and}\, 0$. Here, we consider all the central galaxies in our sample (i.e. with number of nearest neighbors $\geq0$). Shaded areas are defined by the $25$th-$75$th interquartile range of the data at each redshift. We divide our galaxies into the four different types as described in Sec. \ref{sec:star_formation_rates_classification}.}
    \label{fig:merger_statistic_observations}
\end{figure*}

Despite all the previously mentioned differences between observations and simulations, the overall result that the presence of a massive and/or dense central component can suppress the development of strong lopsidedness by stabilizing the disk is consistent both in observations and simulations at high-redshift.

\subsection{Origin of lopsidedness: galaxy interactions vs. gas accretion}
\label{sec:origin_lopsidedness}
Similarly to Fig. \ref{fig:merger_statistic} and \ref{fig:net_gas_accretion}, in the left panel of Fig. \ref{fig:merger_statistic_observations}, we show the fraction of galaxies with one or more massive neighbors (i.e. stellar mass-ratio $>1$:$10$) within $R_{200}$ (left panel), considering only the sub-sample of the central galaxies with at least one nearest neighbor. In the right panel of Fig. \ref{fig:merger_statistic_observations}, we show the net accretion rate onto the galactic disk within the last $1\, \mathrm{Gyr}$, considering all the central galaxies (i.e. with number of nearest neighbors $\geq0$) in our selected samples between $0 < z < 2$. 
We note that, in Fig. \ref{fig:merger_statistic_observations}, we consider only the central galaxies, divided into the four different galaxy types defined in Sec. \ref{sec:star_formation_rates_classification}. 

We see that the fraction of TypeI galaxies with one or more massive neighbors within $R_{200}$ remains overall constant at $40\%$ as a function of redshift. 
Since this fraction is also the highest compared to the other galaxy types, it indicates that close tidal interactions are typically more significant in TypeI galaxies and, thus, more often behind the triggering of their strong lopsidedness at all redshift, also as a result of their internal properties (i.e. low central stellar mass density, low core mass fraction and large disk size).  
However, as previously discussed in Sec. \ref{sec:origin_lopsidedness}, close tidal interactions cannot be the only mechanism at play triggering lopsidedness.
In fact, in the right panel of Fig. \ref{fig:merger_statistic_observations}, we see that TypeI galaxies are always characterized by positive net accretion rate, slightly decreasing towards low-redshift. Specifically, at high-redshift (i.e. $z>1$), we find that TypeI galaxies have experienced significant recent gas accretion with respect to the other galaxy types. This is overall consistent with the results from \citet{LeBail2023}, where they found that their TypeI galaxies are characterized by clumpy and heterogeneous star-forming disks, suggesting a recent gas-rich major merger.
Overall, these results suggest that both close tidal interactions and recent gas accretion can play a fairly relevant role in the origin of lopsidedness at all redshift between $0 < z < 2$. 

On the other hand, the fraction of TypeII galaxies with one or more massive neighbors within $R_{200}$ decreases towards low-redshift, similarly to that of TypeIII galaxies. However, at $z\lesssim0.3$, the fraction of TypeIII galaxies with one or more massive within $R_{200}$ suddenly rises again to $\sim30\%$. 
Furthermore, we see that, while TypeIII galaxies show an overall positive and constant net accretion rate at all redshift, TypeII galaxies show decreasing net accretion rate towards low-redshift. 
Although we do not show it here, we note that we find similar behaviors of the net accretion rate for the different galaxy types when we consider, separately, central galaxies with at least one or no massive neighbors within $R_{200}$ at any chosen redshift. This suggests that a recent gas accretion event onto the disks of our galaxies is not necessarily related to the accretion of gas from nearby satellite galaxies. For example, this gas accretion could be primarily driven from the intergalactic medium. The study of the source of gas accretion onto our galaxies, as well as its potential for producing lopsidedness, will be left to a follow-up work. 
Overall, this suggests that, at the highest redshift (i.e. $z\gtrsim1.5$), close tidal interactions and recent gas accretion are similarly affecting both TypeII and TypeIII galaxies, thus producing a similar lopsided perturbation in these galaxies, as shown in Fig. \ref{fig:asymmetry_vs_galaxy_properties}.
However, the fact that TypeII galaxies are characterized by quenched disks and TypeIII galaxies by star-forming disks suggests either that the gas is being accreted with different angular momentum, such that it rapidly in-falls towards the inner galactic regions in TypeII galaxies, or that different AGN feedback processes are active in TypeII and TypeIII galaxies \citep{Irodotou2022}. Alternatively, the induced lopsidedness could also be responsible of funneling material towards the inner galactic regions due to internal torques. This latter outside-in quenching scenario induced by the presence of a strong lopsided perturbation has been previously proposed by \citet{Kalita2022}, who observed highly star-forming bulges embedded in quiescent lopsided stellar disks in three group galaxies at $z\sim3$. It remains to be explored what mechanisms are primarily driving the quenching of the disk/central in TypeII galaxies with respect to TypeIII galaxies at high-redshift.
At low-redshift (i.e. $z\lesssim1$), recent gas accretion is mainly affecting TypeIII galaxies. The same is true for close tidal interactions, specifically at the lowest redshift (i.e. $z\lesssim0.3$) where we see that they are more significant in TypeIII than TypeII galaxies. 
This suggests that lopsidedness in TypeIII galaxies can also be triggered by both recent gas accretion and close tidal interactions at low-redshift. 
In the case of TypeII galaxies, even though close tidal interactions can still be at play, they seem less likely to produce strong perturbations due to the high central stellar mass density, high core mass fraction and small disk size of these galaxies (see Fig. \ref{fig:asymmetry_vs_galaxy_properties}). For this reason, we find a lower asymmetry in TypeII than in TypeIII galaxies at low-redshift.   

Finally, for TypeIV galaxies, we see that the fraction of galaxies with one or more massive neighbors increases towards low-redshift. Nonetheless, TypeIV galaxies are always characterized by low asymmetry at all redshift. This is likely the result of the specific internal properties of these galaxies (i.e. high central stellar mass density, high core mass fraction and small disk size), which makes them less prone to develop strong lopsidedness due to external interactions. 
Furthermore, we see that TypeIV galaxies are characterized by negative net accretion rate at all redshift, ruling out recent gas accretion as trigger of lopsidedness in these galaxies. In fact, we find that TypeIV galaxies are the most symmetric galaxies both at high- and low-redshift, as shown in Fig. \ref{fig:asymmetry_vs_galaxy_properties}.
We note that the lack of recent gas accretion does not exclude that TypeIV galaxies may have either experienced earlier accretion of gas, which was rapidly converted into stars, or gas-poor mergers that contributed to the rapid mass assembly of these galaxies (i.e. high core mass fraction and high central stellar mass density), specifically at high-redshift.

\section{Summary and conclusions}
\label{sec:conclusions}
In this work, we have characterized lopsidedness in the mass distribution of the stellar component of our samples of disk-like galaxies selected from the TNG50 simulation at specific redshift between $0 < z < 2$. We remind the reader that we are selecting a new galaxy sample at each redshift, using the selection criteria described in Sec. \ref{sec:sample_selection}. Similarly to \citet{Varela-Lavin2023} and \citet{Dolfi2023}, we quantify lopsidedness via Fourier analysis by decomposing the face-on projected galaxy stellar mass distribution in concentric radial annuli and measuring the amplitude of the first Fourier $m=1$ mode in each radial bin within $R_{\mathrm{h}} < r < 1.4\, R_{90}$ and $|h_{z}| < 2\, h_{90}$, as described in Sec. \ref{sec:lopsidedness}. We then quantify the fraction of lopsided galaxies as a function of redshift, and we study the correlation between lopsidedness and both the local environmental density and galaxy internal properties as a function of redshift. Below, we summarize our main results:

\begin{itemize}
    \item We find that the median lopsided amplitude, as well as the fraction of lopsided galaxies, decreases from high- to low-redshift. Additionally, we find that, while both central and satellite galaxies show similar lopsided amplitude at high-redshift (i.e. $z>1$), central galaxies tend to be more lopsided than satellite ones at low-redshift (i.e. $z\lesssim1$). Overall, this suggests that mechanisms triggering lopsidedness are different or more efficient at high- than low-redshift. In particular, environmental interactions could be playing a more active role in triggering lopsidedness at high- than low-redshift. 

    \item We find that, independently of the redshift considered, the lopsided amplitude does not strongly depend on the local environmental density of the galaxies, consistent with the mild or lack of correlation observed between lopsidedness and the local environment at $z=0$ \citep{Wilcots2010,Dolfi2023}. 

    \item We find a strong correlation between the median lopsided amplitude and the internal properties of the galaxies at all redshifts considered up to $z=2$. This suggests that, independent of the mechanisms producing lopsidedness, galaxies with low central stellar mass density, large stellar half-mass radius and large disk size are more prone to develop strong lopsidedness, consistent with previous results at $z=0$ \citep{Reichard2008,Varela-Lavin2023,Dolfi2023}. 
\end{itemize}

In Sec. \ref{sec:mergers} and \ref{sec:gas_accretion}, we study the relative importance as a function of redshift of different mechanisms triggering lopsidedness, specifically focusing on close tidal interactions and gas accretion, to explore whether different mechanisms are at play at high- and low-redshift, or whether the same mechanisms driving lopsidedness at $z\sim0$ are simply more efficient at high-redshift. Below, we summarize our main findings:

\begin{itemize}
    \item As proxy of the probability of close tidal interactions, we quantify the number of nearest neighbors with stellar mass-ratio $>1$:$10$ located within the virial radius of the galaxy $R_{200}$. In Fig. \ref{sec:mergers}, we find that the fraction of lopsided galaxies with at least one massive neighbors within $R_{200}$ decreases towards low-redshift, suggesting that close tidal interactions can be more frequent and, thus, more efficient at triggering lopsidedness at high-redshift (i.e. $z\gtrsim1$). At lower redshift, the role of close tidal interactions becomes less significant. However, we also find that the majority of lopsided galaxies does not have any massive neighbors within $R_{200}$ (i.e. $80\%$ at $z\gtrsim1$ and $\sim90\%$ at $z<0.5$), suggesting that other mechanisms must be at play and be more relevant in the origin of lopsidedness at all redshift. 

    \item We calculate the net accretion rate onto the galactic disk within the last $1\, \mathrm{Gyr}$ for the galaxies at different redshift between $0 < z < 2$, as described in Sec. \ref{sec:gas_accretion}. We find that lopsided galaxies are always characterized by positive and larger net accretion rate compared to symmetric galaxies, which mainly show negative net accretion rate. In particular, lopsided galaxies show significant net accretion rate at high-redshift, decreasing towards $z=0$. This suggests that gas accretion can play a significant role in triggering lopsidedness at all redshift, specifically at high ones (i.e. $z>1$).  
\end{itemize}

In summary, we find that multiple mechanisms (i.e. gas accretion and close tidal interaction) are playing a more active and efficient role in triggering lopsidedness at high-redshift (i.e. $z\gtrsim1$). This would mean that the seemingly more widespread lopsidedness detected with observations at high-redshift \citep{LeBail2023} can be associated to the more efficient conditions existing at high-redshift. 
In a follow-up work, we aim to study the distribution of lopsidedness within the large-scale environment as a function of redshift to determine how the connectivity of galaxies to filaments can influence the distinct net accretion rate, as well as star formation histories of galaxies up to $z=0$ (see figures 13 and 14 in \citet{Dolfi2023}). This will also allow us to understand the origin of lopsidedness through asymmetric gas accretion with subsequent star formation in realistic cosmological models.

In Sec. \ref{sec:comparison_observations}, we perform a more detailed comparison with the observations of \citet{LeBail2023} by classifying our galaxies into four galaxy types depending on the star formation rate of the central and disk components, similarly to \citet{LeBail2023}. The galaxy classification is described in details in Sec. \ref{sec:star_formation_rates_classification}.
Below, we summarize our main results:

\begin{itemize}
    \item We find that, at high-redshift (i.e. $z > 1$), the lopsided amplitude decreases with increasing central stellar mass density from TypeI to TypeII/TypeIII to TypeIV galaxies. This suggests that the central stellar mass density is one of the key parameters responsible for determining the strength of the lopsided perturbations at high-redshift. This is consistent with the more efficient role of external interactions in perturbing the galaxies at high-redshift, which still require specific internal galaxy properties to develop strong lopsidedness. On the other hand, at low-redshift, we find that the disk size of the galaxies is playing a more relevant role for lopsidedness. This is consistent with the finding from Sec. \ref{sec:origin_lopsidedness} that lopsidedness is primarily triggered as a result of gas accretion rather than external interactions at low-redshift.
    
    \item We find that results in Fig. \ref{fig:asymmetry_vs_galaxy_properties} slightly differs from the observational results of \citet{LeBail2023} at high-redshift. 
    Specifically, \citet{LeBail2023} find that TypeIII galaxies are the most symmetric with the highest core mass fraction in their quenched bulge. On the other hand, we find that TypeIII galaxies have similar lopsidedness, but lower core mass fraction, than TypeII galaxies.
    In Sec. \ref{sec:asymmetry}, we note several reasons that could produce this discrepancy between observations and simulations at high-redshift. However, despite these differences, the general result that {\it the presence of a centrally massive/dense bulge component can suppress the development of strong lopsidedness is consistent both in simulations and observations}.
    
    \item We find that close tidal interactions are typically more effective in triggering lopsidedness in TypeI galaxies than in TypeII, TypeIII and TypeIV galaxies at all redshift. Furthermore, we find that TypeI galaxies experience significant recent gas accretion at high-redshift, which could be behind their strong lopsidedness. This is consistent with the observations from \citet{LeBail2023}, where they find that TypeI galaxies have clumpy and heterogeneous star-forming disks that could be signature of a recent gas-rich major merger. 

    \item In Fig. \ref{fig:merger_statistic_observations}, we find that TypeII galaxies experience recent gas accretion at high-redshift, similarly to TypeIII galaxies. The quenching of the disk in TypeII galaxies can then be a result of the specific angular momentum of the accreted gas, causing it to in-fall towards the central regions, or different AGN feedback processes being active in TypeII and TypeIII galaxies \citep{Irodotou2022}. Alternatively, the lopsidedness induced by the recent gas accretion could be responsible for funneling the gas towards the inner galactic regions due to internal torques, leaving a star-forming central region embedded in a quenched lopsided stellar disk at high-redshift. This outside-in quenching scenario induced by lopsidedness has been proposed by \citet{Kalita2022}, who observed strong lopsided quiescent stellar disks and highly star-forming bulges in three group galaxies at $z\sim3$. While this does not exclude that the quenched disks in TypeII galaxies could result from different mechanisms, it shows that lopsidedness could have significant impact for the evolution of galaxies. 
\end{itemize}

In a follow-up work, we aim to investigate in more details the effect that lopsidedness has on galactic disks and inner galactic regions, with the consequences for the future evolutionary histories of the galaxies.

\begin{acknowledgements}
The authors gratefully acknowledge support by the ANID BASAL project FB210003. A.M. acknowledges support from the FONDECYT Regular grant 1212046. F.A.G. acknowledges support from the FONDECYT Regular grant 1211370. A.M. and F.A.G. gratefully acknowledge funding from the Max Planck Society through a “PartnerGroup” grant as well as from the HORIZON-MSCA-2021-SE-01 Research and Innovation Programme under the Marie Sklodowska-Curie grant agreement number 101086388. P.B.T. acknowledges partial funding from FONDECYT 1240465 and the LACEGAL Network (Horizon 2030). 
\end{acknowledgements}

\section*{Data Availability}
The data used in this work come from the IllustrisTNG simulations, which are publicly available here: \url{https://www.tng-project.org/data/}, and described in \citet{Nelson2019}.

%
%
\bibliographystyle{aa}
\bibliography{biblio} 

\end{document}